\documentclass[12pt,preprint]{aastex}

\newcommand{\msun}{M$_\sun$}
\newcommand{\lsun}{L$_\sun$}

\newcommand{\hinv}{$h^{-1}\ $}
\newcommand{\om}{$\Omega_m$}

\newcommand{\sigv}{$\sigma_{v}$}
\newcommand{\ngal}{$N_{gal}$}

\newcommand{\lam}{$\Lambda$}

\slugcomment{Version 7.0, 
Last modified 05-May-2003}

\shorttitle{Bahcall et al.}
\shortauthors{}

\begin{document}


\title{A MERGED CATALOG OF CLUSTERS OF GALAXIES	FROM EARLY SDSS DATA}


\author{
Neta A. Bahcall\altaffilmark{1},
Timothy A. McKay\altaffilmark{2},
James Annis\altaffilmark{3},
Rita S.J. Kim\altaffilmark{4},
Feng Dong\altaffilmark{1},
Sarah Hansen\altaffilmark{2},
Tomo Goto\altaffilmark{5},
James E. Gunn\altaffilmark{1},
Chris Miller\altaffilmark{5},
R. C. Nichol\altaffilmark{5},
Marc Postman\altaffilmark{6},
Don Schneider\altaffilmark{7},
Josh Schroeder\altaffilmark{1},
Wolfgang Voges\altaffilmark{8},
Jon Brinkmann\altaffilmark{9},
Masataka Fukugita\altaffilmark{10}
}

\altaffiltext{1}{Princeton University Observatory, Princeton, NJ 08544}
\altaffiltext{2}{University of Michigan, Department of Physics, 500 East 
University, Ann Arbor, MI 48109}
\altaffiltext{3}{Fermi National Accelerator Laboratory, P.O. Box 500, 
Batavia, IL 60510}
\altaffiltext{4}{Department of Physics and Astronomy, The Johns Hopkins University, 
Baltimore, MD 21218}
\altaffiltext{5}{Dept. of Physics, Carnegie Mellon University, 5000 Forbes 
Ave., Pittsburgh, PA-15232}
\altaffiltext{6}{Space Telescope Science Institute, Baltimore, MD 21218}
\altaffiltext{7}{Department of Astronomy and Astrophysics, The Pennsylvania State University, University 
Park, PA 16802}
\altaffiltext{8}{Max-Planck-Institut f\"{u}r Extraterrestrische Physik, 
D-85740 Garching, Germany}
\altaffiltext{9}{Apache Point Observatory, 2001 Apache Point Road, P.O. Box 59, Sunspot, NM 88349-0059}
\altaffiltext{10}{Institute for Cosmic Ray Research, University of Tokyo, 
Midori, Tanashi, Tokyo 188-8502, Japan}


\begin{abstract}
We present a catalog of 799 clusters of galaxies in the redshift range
z$_{est}$ = 0.05 - 0.3 selected from $\sim$400 deg$^2$ of early SDSS commissioning data along the celestial equator.  
The catalog is based on merging two independent selection methods 
-- a color-magnitude red-sequence maxBCG technique (B), and a Hybrid Matched-Filter method (H). The BH catalog 
includes clusters with richness \lam $\geq$ 40 (Matched-Filter) and \ngal $\geq$ 13 
(maxBCG), corresponding to typical velocity dispersion of 
\sigv $\ga$ 400 km s$^{-1}$ and mass (within 0.6 $h^{-1}$ Mpc radius) $\ga 
5\times10^{13}\ h^{-1}$ \msun. This threshold is below Abell richness class 0 clusters. 
The average space density of these clusters is $2\times10^{-5}\ h^3$ Mpc$^{-3}$. 
All NORAS X-ray clusters and 53 of the 58 Abell clusters in the survey region are detected in the catalog; 
the 5 additional Abell clusters are detected below the BH catalog cuts. 
The cluster richness function is determined and found to exhibit a steeply decreasing cluster abundance 
with increasing richness. We derive observational scaling relations between cluster 
richness and observed cluster luminosity and cluster velocity dispersion; these scaling relations provide 
important physical calibrations for the clusters. The catalog can be used for studies of individual clusters, 
for comparisons with other sources such as X-ray clusters and  AGNs, and, with proper correction for the relevant selection 
functions, also for statistical analyses of clusters.
\end{abstract}




\section{Introduction}
Clusters of galaxies, the largest virialized systems known, provide one of 
the most powerful tools in studying the structure and evolution of the 
Universe. Clusters highlight the large scale structure of the universe 
\citep{abe58, bahn83, bahn84, kly83, bahn88, huc90, pos92, cro97}; 
they trace the evolution of structure with time \citep{hen92,  
eke96, bahn97, car97, bahn98, don99, hen00, ros02}; 
they constrain the amount and distribution of dark and baryonic 
matter \citep{zwi57, abe58, bahn77, whi93, bahn95, fis97, car97, carl01}; 
they reveal important clues about the formation and evolution of galaxies 
\citep{dre84, gun88}; and they place critical constraints 
on cosmology \citep{bahn92, whief93, eke96, car97, bahn98, bahn99}. 
In fact, clusters of galaxies place some of the most powerful constraints 
on cosmological parameters such as the mass density of the 
Universe and the amplitude of mass fluctuations. 
In spite of their great value and their tremendous impact 
on understanding the Universe, systematic studies of clusters of galaxies 
are currently limited by the lack of large area, accurate, 
complete, and objectively  selected catalogs of optical clusters, and by the 
limited photometric and redshift information for those that do exist. 

The first comprehensive catalog of clusters of galaxies ever 
produced, the Abell Catalog of Rich Clusters \citep{abe58, abe89}, was a 
pioneering project that provided a seminal contribution to the study of 
extragalactic astronomy and to the field of clusters of galaxies. While 
galaxy clustering had been recognized before Abell, the data were fragmentary 
and not well understood. Both Abell's catalog, as well as Zwicky's 
\citep{zwi68} independent catalog, were obtained by visual inspection of 
the Palomar Observatory Sky Survey plates. These catalogs have served the 
astronomical 
community for nearly half a century and were the basis for many of the 
important advances in cluster science (see references above; also 
Abell's Centennial paper, \citealt{bahn99b}). At the beginning of the new century, 
the need for a new 
comprehensive catalog of optical clusters -- one that is automated, precise, 
and objectively selected, with redshifts that extend beyond the z$\la$0.2 limit 
of the Abell catalog -- has become apparent. 

There have been recent advances in this direction, 
including large area catalogs selected by objective algorithms from digitized 
photographic plates (\citealt{she85} for the Lick Catalog; \citealt{lum92} 
for the EDCC Catalog; \citealt{dal94} and \citealt{cro97} 
for the APM catalog), as well as small area, deep digital surveys of distant 
clusters (e.g., the 5 deg$^2$ Palomar Distant Cluster Survey, 
\citealt{pos96}; 100 deg$^2$ Red-Sequence Cluster Survey, \citealt{gla00}; 
and 16 deg$^2$ KPNO Deeprange Survey, \citealt{pos02}). 
A particularly important advance for optical surveys has been the inclusion
of accurate CCD-based color information for galaxy selection. The inclusion
of color in cluster selection greatly reduces the problems of density
projection which have long plagued optical selection of clusters. Good
examples of color-based optical selection include the 100 deg$^2$ 
Red-Sequence Cluster Survey \citep{gla00} and the SDSS selection described
in this work.

Surveys of X-ray clusters and observations of the Sunyaev-Zeldovich 
effect in clusters have and will continue to provide important data that is 
complementary to the optical observations of clusters of galaxies. These 
methods identify rich systems that have developed an extensive
hot intracluster medium. While excellent for selection of massive, well 
developed clusters, these methods have thresholds which are sensitive to
the evolution of the hot intracluster medium, both with cosmic time
and with the richness of the objects. In this sense, optical selection 
has the important complementary advantage of
being able to identify galaxy clustering across a wide range of
system richness and time evolution.

The Sloan Digital Sky Survey (SDSS; \citealt{yor00}) will provide a 
comprehensive digital imaging survey of  10$^4$ deg$^2$ of the North 
Galactic Cap (and a smaller, deeper area in the South) in five bands ($u$, $g$, $r$, 
$i$, $z$), followed by a spectroscopic multi-fiber 
survey of the brightest one million galaxies (\S \ref{clusterselection}). With high 
photometric precession in 5 colors and a large area coverage (comparable 
to the Abell catalog), the SDSS survey will enable state-of-the-art cluster 
selection using automated cluster selection methods. 
Nearby clusters (to z $\la$ 0.05 - 0.1) can be selected directly in 3-dimensions using 
redshifts from the spectroscopic survey. The imaging survey will enable 
cluster selection to z$\sim$0.5 and beyond using the 5 color bands of the survey.
In the range z$\sim$0.05 - 0.3, the 2D cluster selection 
algorithms work well, with only small effects due to selection function (for the 
richest clusters). In the nearest part of this range, 
z$\sim$0.05 - 0.15, the SDSS spectroscopic data can also be useful for
cluster confirmation and for redshift determination. Even poor clusters
can be detected with high efficiency in this redshift range. For z$\sim$0.3 - 0.5, 
2D selection works well, but selection function effects become important, 
especially for poorer clusters. 

Several cluster selection algorithms have recently been 
applied to $\sim$400 deg$^2$ of early SDSS imaging 
commissioning data in a test of various 2D cluster selection 
techniques. These methods, outlined in \S \ref{clusterselection}, include the 
Matched-Filter method \citep{pos96, kep99, kim02a}, and the red-sequence color-magnitude 
method, maxBCG \citep{ann02}, as well as a Cut and Enhance method \citep{got02} and a multicolor 
technique (C4; \citealt{mil02}). Each method can identify clusters of galaxies in SDSS data 
to z$\sim$0.5, with richness thresholds that range from small groups 
to rich clusters, and with different selection functions. 
Since each algorithm uses different selection criteria that emphasize different 
aspects of clusters, the lists of clusters found by different techniques will not be identical. 

In this paper we present a catalog of 799 clusters of galaxies in the redshift range z = 0.05 - 0.3 
from 379 deg$^2$ of SDSS imaging data. The catalog was constructed by merging lists of clusters 
found by two independent 2D cluster selection methods: Hybrid Matched Filter and maxBCG. We compare the 
results from the two techniques and investigate the nature of clusters they select. We derive scaling 
relations between cluster richness and observed cluster 
luminosity and cluster velocity dispersion. We use the scaling relations to 
combine appropriate subsamples of these lists to produce a conservative 
merged catalog; the catalog is limited to a richness threshold specified in \S \ref{catalog}; 
the threshold corresponds to clusters with a typical velocity dispersion of 
$\sigma_{v}\ga$ 400 km s$^{-1}$. The average space density of the clusters is 
$\sim 2\times10^{-5} h^{3}$ Mpc$^{-3}$. A flat LCDM cosmology with \om = 0.3 and a Hubble 
constant of H$_0$ = 100 $h$ km s$^{-1}$ Mpc$^{-1}$ with $h$ = 1 is used throughout. 
The current work represents preliminary tests of selection algorithms on early SDSS commissioning data. 
The results will improve as more extensive SDSS data become available.

\section{Cluster Selection from SDSS Commissioning Data} \label{clusterselection}

The SDSS imaging survey is carried out in drift-scan mode in five filters, $u$, $g$, $r$, $i$, $z$, 
to a limiting magnitude of $r<$23 \citep{fuk96, gun98, lup01, hog01}. 
The spectroscopic survey will target nearly one million 
galaxies to approximately $r<$17.7, with a median redshift of z$\sim$0.1 
\citep{str02}, and a small, deeper sample of $\sim$10$^5$ Luminous Red 
Galaxies to $r\sim$19 and z$\sim$0.5 \citep{eis01}. For 
more details of the SDSS survey see \citet{yor00}, \citet{bla02}, \citet{pie02}, \citet{smi02} and \citet{sto02}. 

Cluster selection was performed on 379 deg$^2$ of SDSS commissioning 
data, covering the area $\alpha$(2000) = 355\degr\ to 56\degr, $\delta$(2000) 
= -1.25\degr\ to 1.25\degr; and $\alpha$(2000) = 145.3\degr\ to 236.0\degr, $\delta$(2000)= 
-1.25\degr\ to 1.25\degr\ (runs 94/125 and 752/756). The limiting magnitude of galaxies used 
in the cluster selection algorithms was conservatively selected to be $r<$21 
(where $r$ is the SDSS Petrosian magnitude). At this magnitude limit, star-galaxy 
separation is excellent \citep{scr01}. The clusters of galaxies studied in this paper were selected from these 
imaging data using a Matched Filter method \citep{kim02a, kim02b} 
and an independent color-magnitude maximum-likelihood Brightest Cluster Galaxy method 
(maxBCG; \citealt{ann02}). These methods are briefly described below.

The Matched Filter method HMF (Hybrid Matched Filter; \citealt{kim02a}) is a Hybrid of the Matched Filter 
\citep{pos96} and the Adaptive Matched Filter techniques \citep{kep99}.  This 
method identifies clusters in imaging data by finding 
peaks in a cluster likelihood map generated by convolving the galaxy survey 
with a filter based on a model of the cluster and field galaxy distributions. 
The cluster filter is composed of a projected density profile model for the 
galaxy distribution (a Plummer law profile is used here), and a luminosity function filter 
(Schechter function). The filters use the typical parameters observed for 
galaxy clusters (e.g., core radius R$_c$ = 0.1 $h^{-1}$ Mpc, cutoff radius 
R$_{max}$=1 \hinv Mpc, and luminosity function parameters M$^*_r=-20.93$ and 
$\alpha=-1.1$ for $h$ = 1). The HMF method identifies the highest likelihood 
clusters in the imaging data ($r$-band) and determines their estimated redshift (z$_{est}$) 
and richness (\lam); the richness \lam \ is derived from the best-fit cluster model 
that satisfies L$_{cl}$($<1\ h^{-1}$ Mpc) = \lam L$^{*}$, where L$_{cl}$ is 
the total cluster luminosity within 1 $h^{-1}$ Mpc radius (at z$_{est}$), and 
L$^{*} \sim 10^{10} h^{-2} L_{\sun}$. A relatively high threshold has been applied to the 
cluster selection ($\sigma>$5.2, \citealt{kim02a}); the selected 
clusters have richnesses \lam $\ga$ 20 - 30 
(i.e., L$_{cl}$($<1 h^{-1}$ Mpc) $\ga 2\times10^{11} h^{-2} L_{\sun}$).
This threshold is below the typical Abell richness class 0.

The maxBCG method (\citealt{ann02}) is based on the fact that the 
brightest cluster galaxy (BCG) generally lies in a narrowly defined space in 
luminosity and color (see, e.g, \citealt{hoe85, gla00}). For each SDSS galaxy, 
a ``BCG likelihood'' is calculated based on the galaxy color ($g-r$ and $r-i$) and 
magnitude (M$_i$, in $i$-band). The BCG likelihood is then weighted by the number of nearby galaxies 
located within the color-magnitude region of the appropriate E/S0 ridgeline; this count includes 
all galaxies within 1 $h^{-1}$ Mpc projected separation that are fainter than M$_i$ and brighter 
than the magnitude limit M$_i$(lim) = -20.25, and are located within 2-$\sigma$ of the mean observed 
color scatter around the E/S0 ridgeline (i.e., $\pm^{0.1^m}_{0.15^m}$). 
The combined likelihood is used for cluster identification. The likelihood is 
calculated for every redshift from z = 0 to 0.5, at 0.01 intervals; the redshift that 
maximizes the cluster likelihood is adopted as the cluster redshift. 
Since BCG and elliptical galaxies in the red ridgeline possess very specific colors and luminosities, their observed magnitude 
and colors provide excellent photometric redshift estimates for the parent clusters. The richness 
estimator, \ngal, is defined as the number of red E/S0 ridgeline galaxies (within the 2-$\sigma$ 
color scatter as discussed above) that are brighter than 
M$_i$(lim) = -20.25 (i.e., 1 mag fainter than L$^*$ in the $i$-band; $h$ = 1), and are located within 
a 1 $h^{-1}$ Mpc projected radius of the BCG. 

The HMF and maxBCG methods focus on different properties of galaxy clusters: HMF finds 
clusters with approximately Plummer density profiles and a Schechter luminosity function, 
while maxBCG selects groups and clusters dominated by red $\sim L^{*}$ galaxies. We compare 
the results of these cluster selection algorithms in the following sections and 
merge the clusters into a single complementary self-consistent catalog.

\section{Comparison of the HMF and maxBCG Clusters} \label{clustercomparison}

When comparing different catalogs, uncertainties in cluster estimated redshift, 
position, richness, and selection function,
in addition to the different nature of each cluster selection 
algorithm, render the comparisons difficult. Even selecting the
richest clusters from each catalog will not provide a perfect match,
mostly due to the noisy estimate of richness and its sharp threshold.
In this section we briefly summarize the main comparisons of 
the cluster redshift, position, and richness estimators for 
the HMF and maxBCG methods. 

The accuracy of cluster redshift estimates for each method was determined using 
comparisons with measured redshifts from the SDSS spectroscopic data. A comparison 
of the estimated and spectroscopic redshifts for HMF and maxBCG clusters
with z$_{est}$ = 0.05 - 0.3 and richnesses \lam $\geq$ 
40 (HMF) and \ngal $\geq$ 13 (maxBCG) is shown in Figures 1 and 2. A spectroscopic 
match is considered if the spectroscopic galaxy is located at the position of the BCG. For these 
relatively high richness clusters we find a redshift uncertainty of $\sigma_{z}$ = 0.014 
for maxBCG (from 382 cluster matches) and $\sigma_{z}$ = 0.033 for HMF (from 237 cluster matches; 
there are fewer HMF matches since a spectroscopic match is defined at the BCG position so as 
to minimize noise). A direct comparison between the HMF and maxBCG 
estimated cluster redshifts, using a positional matching criterion defined below, is shown in 
Figure 3.

The positional accuracy of cluster centers is determined by comparing HMF-maxBCG cluster pairs 
(in the above z = 0.05 - 0.3 sample) with pairs in random catalogs. The comparison shows significant excess 
of cluster matches over random for projected cluster separations of $\la$ 0.5 $h^{-1}$ Mpc, with a 
tail to $\sim$ 1 $h^{-1}$ Mpc (Figure 4). 
These excess pairs represent real cluster matches; their distribution provides a measure of the typical offset 
between the cluster centers determined in the two methods. The offsets follow a Gaussian 
distribution with a dispersion of 0.175 $h^{-1}$ Mpc (Figure 4).

Comparison of clusters identified by different selection methods depends not only on the 
positional and redshift uncertainties discussed above, and on the different selection function 
inherent to each catalog, but also on the uncertainties in the richness estimates. The difference in 
selection functions and the uncertainties in richness estimates are the main cause of the relatively 
low matching rates among different samples (see \S \ref{catalog}). The richness scatter is important because 
each cluster sample is cut at a specific richness threshold; since the observed richness function 
is steep and the richness scatter is significant, a richness threshold causes many 
clusters to scatter across the threshold.  This scatter has 
a strong effect on cluster sample comparisons. We illustrate the effect by Monte Carlo simulations 
of two identical cluster samples with different noisy richness estimators (Figure 5). Placing a richness 
threshold on each sample, we obtain richness limited subsamples. For an intrinsic richness function of 
$N_{cl}\propto$(richness)$^{-4}$ (see \S \ref{abundance}), and a 30 $\%$ scatter in richness, the overlap of the two samples is 
only 54 $\%$. Any difference in selection functions, which can be nearly a factor of 
$\sim$2 in the two methods used here, will further reduce the apparent overlap. This simple
model provides an estimate for how large we might expect the overlap between two otherwise 
identical cluster samples to be. It is important to bear this in mind as we make direct comparisons of 
cluster catalogs in subsequent sections.
 
How do the HMF and maxBCG cluster richness estimates compare with each other? 
Cluster richness estimates describe, in one form or another, how 
populated or luminous  a cluster is: either by counting galaxy members 
within a given 
radius and luminosity range, or by estimating total cluster luminosity. 
In general, this measure also reflects the mass of the cluster, its 
velocity dispersion, and temperature. While richness 
correlates well on average with other parameters (e.g., rich clusters 
are more luminous and more massive than poor clusters),
individual cluster richness estimates exhibit large scatter. 
This scatter is due to the sharp luminosity threshold in the richness galaxy 
count, uncertainties in the background corrections, 
uncertainties in the estimated redshift and center of the cluster, 
sub-structure in clusters, and other effects. Still, optical richness 
estimators provide a basic measure of a cluster population; richnesses 
have been determined for all clusters in the above catalogs. The two richness estimates 
obtained by the cluster selection algorithms described above are \ngal\ for maxBCG 
and \lam \ for HMF (\S \ref{clusterselection}). \ngal\ is the number of red (E/S0) ridgeline 
galaxies located within 1 \hinv Mpc of the BCG galaxy and are brighter then M$_i$(lim) = -20.25. 
The richness \lam\ is determined by the HMF fine likelihood for each cluster and reflects 
the best-fit cluster model luminosity within 1 \hinv Mpc radius, L$_{cl}$($<1 h^{-1}$ Mpc)= \lam\ L$^*$ 
(\S \ref{clusterselection}; see \citealt{kep99} and \citealt{kim02a}). In comparing these richnesses, 
differences in the estimated redshifts and cluster centers 
introduce additional scatter on top of any intrinsic variations. 

Figure 6 presents the observed relation between \lam\ and \ngal\ for clusters with z$_{est}$ = 0.05 - 0.3 and 
\ngal $\geq$ 13. While the scatter is large, as expected from the Monte Carlo 
simulations (Figure 5), a clear correlation between the mean richnesses is observed. The best-fit 
relation between \ngal\ (as determined for the maxBCG clusters with \ngal$\geq$13) and the mean 
\lam\ (for the matching HMF clusters) is: 
\begin{eqnarray}
\Lambda = (11.1\pm0.8)\ N_{gal}^{(0.50\pm0.03)}  
\end{eqnarray}
The error-bars reflect uncertainties on the mean best-fit.
(This relation differs somewhat if both richnesses are determined at the maxBCG-selected cluster positions and 
redshifts or at the HMF-selected clusters; see, e.g., \citealt{ann02b}). 
The ratio \lam/\ngal\ decreases somewhat with \ngal; we find \lam/\ngal\ 
$\simeq$ 2 for \ngal$\ga$20, increasing to \lam/\ngal\ $\sim$ 3 for lower richnesses.

A comparison of the richness estimates \lam\ and \ngal\ with directly observed cluster 
luminosities and velocity dispersions is discussed in the following section.

\section{Cluster Scaling Relations: Richness, Luminosity, and Velocity Dispersion}
\label{scaling}

We derive preliminary scaling relations between cluster richness estimates 
and directly observed mean cluster luminosity and cluster velocity dispersion. This enables a direct 
physical comparison between the independent catalogs and allows proper merging of the two samples. 
It also provides a physical calibration of the cluster richness estimates in terms of their mean 
luminosity, velocity dispersion, and hence mass.

\subsection{Cluster Luminosity} \label{lumscale}

The observed cluster luminosities can be directly 
obtained from the SDSS imaging data using population subtraction. By comparing the galaxy population
in regions around cluster centers to that in random locations we can
determine the properties of galaxies in and around the clusters as well as the cluster luminosities. 
Since the redshifts of the SDSS clusters 
are relatively accurate, we can determine cluster luminosities in physical units --- i.e., in 
solar luminosities within a metric aperture. The multi-color SDSS data
also allow us to apply accurate k-corrections to cluster galaxy magnitudes.

We determine the luminosity of a cluster by measuring the total luminosity
of all galaxies within 0.6 \hinv Mpc of the cluster center. We use all HMF and maxBCG clusters 
in the redshift range 0.05 $\leq$ z $\leq$ 0.3 with richness \lam $\geq$ 30 (HMF) and \ngal $\geq$ 10 
(maxBCG). For each cluster, we extract all galaxies within a projected radius of 0.6 \hinv Mpc of 
the cluster center, and compute a k-corrected absolute magnitude for each galaxy according to its 
type (following \citealt{fuk96}). We then sum the total luminosity ($r$-band) within the absolute 
magnitude range of -23.0$\leq M_{r}\leq$-19.8. We determine the background contribution
to this total luminosity by selecting five random locations away from the cluster area (within the same 
SDSS stripe), each with the same angular extent; we extract galaxies within these regions,
k-correct them as if they were at the cluster redshift, and subtract the resulting mean luminosity 
(within the same magnitude range) from that of the cluster. This process allows a determination of 
the variance in the background correction and yields an estimate of cluster 
luminosity within a radius of 0.6 \hinv Mpc and within 
the luminosity range -23.0$\leq M_{r}\leq$-19.8 (corresponding to approximately 1.3 mag below 
HMF's $L^*_r$). We denote this luminosity $L^{{}^r}_{0.6}$. A Hubble constant of $h$ = 1 and 
a flat LCDM cosmology with \om = 0.3 are used to determine cluster distances and luminosities. 
Details of this analysis, along with tests and a variety of related population subtraction results, 
will be presented in a forthcoming paper \citep{han03}.

For greater accuracy, and to minimize the spread due to redshift uncertainty, all clusters with a given 
richness are stacked and their mean luminosity $L^{{}^r}_{0.6}$ determined. These stacked luminosities are presented as 
a function of cluster richness in Figures 7 and 8 for the HMF and maxBCG clusters. A strong correlation between 
richness and mean luminosity is observed; this is of course expected, since 
both \ngal\ and \lam\ represent cluster richnesses which broadly relate to luminosity (\S 
\ref{clustercomparison}). The best-fit power-law relations to the binned mean luminosities are:
\begin{eqnarray}
L^{{}^r}_{0.6} (10^{10} L_\sun) = (1.6\pm0.4)\ N_{gal}^{1\pm0.07}  &  (maxBCG;  N_{gal} \simeq 10-33) \\  
L^{{}^r}_{0.6} (10^{10} L_\sun) = (0.013\pm0.004)\ \Lambda^{1.98\pm0.08} &    (HMF;  \Lambda \simeq 30-80)
\end{eqnarray}
The few highest richness points (\lam $>$ 80, \ngal $>$ 33) exhibit large scatter due to their small numbers. 
Inclusion of these points does not change the fits; we find $L^{{}^r}_{0.6}$ = 1.6 \ngal\ 
(for maxBCG, \ngal$\geq$10) and $L^{{}^r}_{0.6}$ = 
0.015 \lam$^{1.95}$ (for HMF, \lam$\geq$ 30). The non-linearity observed in the $L$-\lam\ relation 
at high \lam\ reflects the fact that the measured cluster luminosity $L$ corrects for an underestimate in 
\lam \ at high richness seen in simulations \citep{kim02a}; the luminosity $L$ measures the true cluster luminosity, 
independent of any uncertainty in cluster richness estimates.

The luminosity $L^{{}^r}_{0.6}$ is the cluster luminosity down to a magnitude of -19.8. To convert this 
luminosity to a total cluster luminosity, we integrate the cluster luminosity function from -19.8$^m$ down 
to the 
faintest luminosities. The luminosity function of HMF clusters (within R = 0.6 $h^{-1}$ Mpc) is observed to have 
Schechter function parameters of $\alpha = -1.08 \pm 0.01$ and M$^{*}_{r} = - 21.1 \pm 0.02$, and maxBCG has 
$\alpha = -1.05 \pm 0.01$ and M$^{*}_{r} = - 21.25 \pm 0.02$ ($h$ = 1; \citealt{han03}). Integrating these 
luminosity functions from -19.8 down to zero luminosity yields correction factors of 1.42 (for HMF) and 1.34 (for maxBCG) for 
the added contribution of faint galaxies to the total cluster luminosity. The total mean cluster luminosities are therefore 
given by Equations 2 and 3 multiplied by these correction factors, yielding 
\begin{eqnarray}
L^{{}^{r,tot}}_{0.6} (10^{10} L_\sun) = (2.1\pm0.5)\ N_{gal}^{1\pm0.07}  &  (maxBCG) \\  
\ \ \ \ L^{{}^{r,tot}}_{0.6} (10^{10} L_\sun) = (0.018\pm0.005)\ \Lambda^{1.98\pm0.08} &    (HMF)
\end{eqnarray}

\subsection{Velocity Dispersion} \label{specscale}

The SDSS spectroscopic survey includes spectra of galaxies brighter
than $r$ = 17.7 \citep{str02}, with a median redshift
of z = 0.1, as well as spectra of the `luminous red galaxy' (LRG) sample
that reaches to r $\simeq$ 19 and z $\sim$ 0.5 \citep{eis01}. For some rich clusters
at low redshift, it is possible within the SDSS spectroscopic data to
directly measure the cluster velocity dispersion. Here 
we compare these velocity dispersions, together with velocity dispersions available from 
the literature (for some of the Abell clusters within the current sample; \S \ref{abell}), 
to cluster richnesses; this provides an independent physical calibration of richness.

The correlation between the observed cluster velocity dispersion and cluster
richness is presented in Figure 9. We use cluster velocity dispersions of 20 clusters determined 
from the SDSS spectroscopic survey (for clusters with $\sim$30 to 160 redshifts) using a Gaussian fit 
method, as well as from several Abell clusters available in the literature (Abell 168, 295, 957, 
1238, 1367, 2644; \citealt{maz96, sli98}). Even though the number of clusters with 
measured velocity dispersion is not large 
and the scatter is considerable, a clear correlation between median velocity
dispersion and richness is observed, as expected (Figure 9). The best-fit relations are:
\begin{eqnarray}
\sigma_{v} (km/s) = (10.2\pm^{13}_6) \ \Lambda^{1\pm0.2} & (HMF;  \Lambda \simeq 30-70) \\ 
\sigma_{v} (km/s) = (93\pm^{45}_{30}) \ N_{gal}^{0.56\pm0.14} & (maxBCG;  N_{gal} \simeq 8-40) 
\end{eqnarray}

Also shown in Figure 9, for comparison, are all stacked SDSS spectroscopic
data for the galaxy velocity differences in the clusters (relative to the
BCG velocity), subtracted for the mean observed background, as a function of 
richness. These are obtained using the best
Gaussian fit to the stacked velocity data, after background subtraction. 
The results are consistent with the directly observed $\sigma$-\lam\ and
$\sigma$-\ngal\ relations discussed above.

The velocity scaling relations (Equations 6 and 7) provide an important calibration of cluster richness 
versus mean cluster velocity dispersion (and thus mass). Also shown in the figures,
for comparison, are the \sigv-richness relations derived from the observed
mean $L_{0.6}$-\lam\ and $L_{0.6}$-\ngal\ correlations (Section \S \ref{lumscale}, 
Figures 7 - 8). Here the luminosity $L^{{}^{tot}}_{0.6}$ is converted to mass, M$_{0.6}$, using the typical
observed M/L ratio relevant for these clusters and the observed relation between
$M_{0.6}$ and \sigv\ based on calibration using gravitational lensing observations 
(see \citealt{bahn03} for details). 
Good agreement exists between these independent scaling relations.
Larger samples, when available, will further improve this important calibration.

\subsection{Consistency of Scaling Relations} \label{consistency}

The independent scaling relations discussed above are consistent with each other.
The directly observed mean \lam-\ngal\ relation (Equation 1) is in agreement
with the observed luminosity-richness relations, $L^{{}^{r,tot}}_{0.6}$-\lam\ and
$L^{{}^{r,tot}}_{0.6}$-\ngal\ (Equations 4 and 5). Both relations ---
the luminosity-richness relations and the \lam-\ngal\ relation --- 
yield, independently, \lam $\simeq 11\ N_{gal}^{0.5}$, and reproduce the observed 
total luminosity relations discussed above. This consistency is illustrated by the
solid and dashed lines in Figure 6 which represent, respectively,
the observed mean \lam-\ngal\ relation and the one obtained from the mean luminosty-richness
relations ($L_{0.6}^{{}^{r,tot}}$-\lam\ and $L_{0.6}^{{}^{r,tot}}$-\ngal).

The third independent relation, velocity dispersion versus richness
(Equations 6 and 7), is also consistent with the above results; this is illustrated
by the dotted curve in Figure 6. 

The non-linearity observed in the $L$(\lam) $\sim$ \lam$^2$ 
relation (Equation 3 and discussion below it; Figure 7), and the similar 
non-linearity observed in the \lam $\sim$ \ngal$^{0.5}$ relation (i.e., \ngal $\sim$ \lam$^2$; 
Equation 1; Figure 6), are consistent with the velocity 
scaling relation, $\sigma$ = 10.2 \lam\ (Equation 6), since the latter implies that 
cluster mass (within a fixed radius) is M $\sim \sigma^2 \sim \Lambda^2$; this is consistent with 
the observed $L \sim \Lambda^2$. The maxBCG relations are 
also self-consistent, with a linear $L \sim$ \ngal, $\sigma \sim$ \ngal$^{0.56}$, 
and hence M $\sim \sigma^2 \sim N_{gal}^{1.1}$. In both cases, $M/L$ is nearly constant --- 
in fact, slightly increasing with $L$ as expected (e.g., \citealt{bahn00}).

The consistency of the scaling relations is illustrated in Figure 6.
A summary of the mean quantitative scaling relations betweem \lam, \ngal,
velocity dispersion, luminosity, and mass (within 0.6 $h^{-1}$ Mpc) is presented
in Table 1.

\section{A Merged Cluster Catalog} \label{catalog}

We use the scaling relations derived above (\S \ref{scaling}) to define a 
conservative merged catalog of clusters of galaxies from the 
early SDSS commissioning data based on the maxBCG and the Hybrid Matched-Filter samples. 
The merged BH catalog is limited to clusters within the redshift range z$_{est}$ = 0.05 - 0.3 
and richness above the threshold listed below, over the 379 deg$^{2}$ area (\S \ref{clusterselection}). 
A total of 799 clusters are listed in the catalog.

The clusters are selected using the following criteria:

\begin{enumerate}
\item z$_{est}$ = 0.05 - 0.3 
\item Richness threshold of \lam $\geq$40 (for HMF clusters) 
and \ngal $\geq$13 (for maxBCG clusters). These thresholds are comparable to each other 
and correspond to a mean cluster velocity dispersion of $\sigma_r\ga$ 400 km s$^{-1}$ and 
luminosity L$^{{}^{r,tot}}_{0.6}\ga 3\times10^{11}\ h^{-2}$ L$_{\odot}$; the related mass is 
approximately M$_{0.6}\ga5\times10^{13}\ h^{-1}$ \msun (see Table 1). 
\end{enumerate}

Clusters that overlap between the two methods are considered as single clusters if they are separated 
by $\leq$1 \hinv Mpc (projected) and $\leq$0.08 in estimated redshift (2.5-$\sigma_z$). 
Overlap clusters are listed as a single cluster, on a single line, but include 
the relevant parameters from both the HMF and maxBCG selection (position, 
redshift, richness). This is done in order to provide complete information about the clusters 
and allow their proper use with the independent HMF and maxBCG selection functions. 
For each cataloged cluster (HMF with \lam$\geq$ 40 or maxBCG with \ngal$\geq$ 13) we include cluster matches 
(i.e., overlaps with separations as defined above) that reach beyond the richness or redshift thresholds of the 
catalog. For example, an HMF cluster with \lam$\geq$ 40 and z = 0.30 may list as a match a maxBCG 
cluster with \ngal$<$ 13 and/or z = 0.22 to 0.38 (i.e., $\Delta$z $\leq$ 0.08). A lower limit of \ngal$\geq$6 
is set for all 
matches. While not part of the \lam$\geq$40, \ngal$\geq$13 catalog, such matches with \ngal$<$13 and \lam$<$40 
clusters are listed in order to provide full information of possible matches, considering the large uncertainty 
in the richness parameter. 
(If there is more than one match per cluster, we select the one with the closest separation).
Some of the matches, especially at low richness (\ngal$\la$10) and large separation ($\sim$ 1 $h^{-1}$ Mpc 
or $\Delta$z $\sim$ 0.08), may be coincidental. 
Clusters that do not overlap are listed as separate clusters and are so noted. 

The catalog is presented in Table 2. Listed in the catalog, in order of increasing right-ascension, are the following: 
SDSS cluster number (column 1), method of detection (H for HMF, B for maxBCG; lower case (h, b) represents 
cluster matches that are outside the catalog richness or redshift thresholds, i.e., \lam$<$40, \ngal$<$13, 
z$>$0.3; column 2), HMF $\alpha$ and $\delta$ (in degrees 2000; column 3 - 4), 
HMF estimated redshift (column 5), HMF cluster richness \lam\ (column 6). Columns 7 - 10 provide similar 
information for the maxBCG detection, if the cluster so detected: $\alpha$ and $\delta$ (2000; column 7 - 8), maxBCG redshift 
estimate (column 9), and richness estimate \ngal \ (column 10). An SDSS spectroscopic redshift that matches the 
cluster, if available, is listed in column 11 (mainly for the BCG galaxy). 
Column 12 lists matches with Abell and X-ray clusters. All the NORAS X-ray clusters and 53 of the 58 Abell 
clusters in this area are identified in the catalog; the additional five Abell clusters 
are identified by the combined HMF and maxBCG techniques but are below the catalog richness threshold 
(see \S \ref{abell}). 

The catalog contains 436 HMF clusters (\lam $\geq$ 40), 524 maxBCG clusters
(\ngal $\geq$ 13), and a total merged catalog (as defined above) of 799 clusters (at $z_{est}$ = 0.05 - 0.3).
Some clusters are false-positive detections (i.e., not real clusters); 
the false-positive rate is discussed below. The overlap between the independent HMF and
maxBCG clusters within the above redshift range is 81$\%$ (of the 
HMF clusters, accounting for all matches to \ngal$\geq$6). This overlap rate is consistent 
with expectations based on the selection functions and false-positive rates 
for the HMF and maxBCG clusters (see below) and the effects of 
redshift and positional uncertainties. The overlap rate increases to $\ga$ 90$\%$ with more liberal matching criteria 
(e.g., separation larger than 1 $h^{-1}$ Mpc and/or larger than 0.08 in redshift). The overlap rate drops, 
as expected, when the richness restriction 
of the matching sample is tightened (e.g., the matching rate is 37$\%$ if only \ngal$\geq$13 matches 
are considered for \lam$\geq$40 HMF clusters; this is consistent with expectations based on Monte Carlo 
richness simulations, \S \ref{clustercomparison}). The richest clusters, HMF with \lam $\geq$52, are matched 
at a higher rate, as expected: 90$\%$ match with \ngal$\geq$6 maxBCG clusters and 61$\%$ match with \ngal$\geq$13 clusters. 
A summary of the catalog cluster distribution by redshift and richness is presented in Table 3.

Selection functions for the independent HMF and maxBCG 
clusters have been determined from simulations and are presented as a function of redshift 
and richness in Figure 10 (for HMF; \citealt{kim02a}) and Figure 11 (for maxBCG; \citealt{ann02}) 
(see above refereneces for more details). The richest clusters are 
nearly complete and volume limited to z $\la$ 0.3, while the 
\lam $\sim$ 40 HMF clusters are only $\sim$40$\%$ complete at z $\sim$ 0.3. 
The selection functions need to be properly accounted for in any statistical analysis of the current samples.

Some systems are false-positive detections (i.e., non-real clusters). The false-positive detection rates for the clusters have been 
estimated from simulations \citep{kim02a, ann02} as well as from visual inspection. The false-positive rate 
is found to be small ($\la$10$\%$) for the \ngal$\geq$13 maxBCG and 
\lam$\geq$40 HMF clusters (z = 0.05 - 0.3). All detections are included in the catalog, including 
false-positive detections,  in order to avoid unquantitative visual selection.

Some maxBCG systems are found to be small clumps of red galaxies in the outskirts of richer HMF clusters. 
Some un-matched HMF and maxBCG systems are in fact parts of the same larger cluster split into separate 
listings because of the $\Delta$z $\leq$0.08 and the 1 $h^{-1}$ Mpc separation cutoff. This can result from 
uncertainties in z$_{est}$ and from the different definitions of cluster center (i.e., HMF clusters typically center on a mean 
high density region, while maxBCG clusters center on a likely BCG galaxy). The splittings may also represent 
sub-structure in clusters. Occasionally, a single HMF or maxBCG cluster may be split by the selection algorithm into two 
separate systems, which may represent sub-clustering. Some systems may be part of 
an extended galaxy overdensity region rather than true condensed virialized clusters; this is less likely for the 
richer systems.

The scaling relations between richness, luminosity and velocity dispersion (\S \ref{scaling}) 
suggest that \lam $\ga$ 40 and \ngal $\ga$ 13 clusters correspond to approximately \sigv $\ga$ 
400 km s$^{-1}$, and \lam $\ga$ 60 and \ngal $\ga$ 30 clusters correspond to \sigv $\ga$ 600 km\ s$^{-1}$, 
i.e., rich clusters. The mean calibrations are summarized in Table 1. 

The distribution of clusters on the sky is mapped for the catalog clusters in Figure 12. 
All clusters with 0.05 $\leq$ z $\leq$ 0.3, richness \lam $\geq$ 40 (for HMF) and 
\ngal $\geq$ 13 (for maxBCG), and their matching clusters are shown. 
The Abell clusters located in the survey area are also 
shown (see \S \ref{abell}). A 1 \hinv Mpc radius circle 
is presented around the center of each cluster; this helps visualize 
possible matches that may be offset in their center position due to 
uncertainties in cluster centers and the different definition of 
``center'' (\S \ref{clustercomparison}), or may represent sub-structure within more extended regions. 

Images of a sample of cataloged clusters representing a wide range of redshift (z$\simeq$0.05-0.3) 
and richness (\lam$\ga$40, \ngal$\ga$13) are presented as examples in Figure 13.

\section{Comparison with Abell and X-Ray clusters} \label{abell}

A total of 58 Abell clusters \citep{abe58, abe89} are located in the current survey region. 
The SDSS BH catalog includes 53 (91$\%$) of these clusters (listed in the last column of Table 2), 
using the matching requirement of a projected separation of less than 1 $h^{-1}$ Mpc. (Since many of 
Abell clusters have no measured redshifts, no redshift information is used.) Most matches are at separations 
typically $\la$ 0.2 $h^{-1}$. The five additional Abell clusters not listed in the catalog are all detected 
by the combined HMF and maxBCG methods, but are below the catalog threshold; these are A116 (\lam = 29, 
\ngal = 9), A237 (\lam = 35, \ngal = 7), A295 (\ngal = 11), A2051 (\ngal = 11), A2696 (\ngal = 11). 
This matching rate is consistent with the expected selection function of the HMF and maxBCG methods. 

Eight clusters from the NORAS X-ray cluster catalog \citep{boh00} lie in the SDSS BH area and redshift 
region. All eight X-ray clusters are detected and included in our catalog; maxBCG detects all eight clusters 
(with 2 below the threshold of \ngal = 13), and HMF detects seven of the clusters (all within the catalog 
threshold of \lam $\geq$ 40). Details of the comparison are given in Table 4.

\section{Cluster Abundance and Richness Function} \label{abundance}

The observed distribution of cluster abundance as a function of richness --- the cluster richness function --- 
is presented in Figure 14. The observed cluster counts are corrected for the 
relevant HMF and maxBCG 
selection functions. Here each cluster is corrected by the selection function appropriate for its richness and 
redshift (for each method; see Figures 10, 11) and by the false-positive expectation rate (\S \ref{catalog}). 
The corrected count is divided by the sample volume to 
produce a volume-limited cluster abundance as a 
function of richness. Smaller corrections for richness and redshift 
uncertainties are not included; these will reduce the cluster abundances by $\sim10\%$ to $\sim30\%$ for 
\lam$\sim$40 to $\sim$60 (see \citealt{bahn03}). 

The results show a steeply declining richness function with increasing richness, as expected. The richness 
function of the HMF-selected and maxBCG-selected clusters are consistent with 
each other when properly corrected for the different selection functions and scaled by the richness scaling 
relation. The richness function indicates a cluster abundance of $2\times10^{-5}\ h^3$ Mpc$^{-3}$ for 
\lam$\ga$ 40 and \ngal$\ga$ 13 clusters ($\sigma\ga$ 400 $km \ s^{-1}$). These abundances are in general good 
agreement with Abell clusters and with other richness or temperature function observations when properly scaled by 
the relevant richness scaling relations (e.g., \citealt{bahn92, ike02}).

The mass function of SDSS clusters was recently determined by \citet{bahn03} (for z = 0.1-0.2, using an extension of 
the current catalog to slightly lower richnesses), yielding consistent results for the HMF and maxBCG subsamples. 
The mass function was used by \citet{bahn03} to place strong 
cosmological constraints on the mass density parameter of the universe, \om, and the amplitude of mass fluctuations, 
$\sigma_8$: \om = 0.19 $\pm^{0.08}_{0.07}$ and $\sigma_8$ = 0.9 $\pm^{0.3}_{0.2}$.

\section{Summary}
We compare two independent cluster selection methods used on 379 deg$^2$ of
early SDSS commissioning data: Matched-Filter (HMF) and the color-magnitude maxBCG.
We clarify the relation between the methods and the nature of clusters they select.
HMF selects clusters that follow a typical density profile and
luminosity function, while maxBCG selects clusters dominated by bright red galaxies ---
quite different selection criteria. We determine scaling relations between the 
observed cluster richness, luminosity, and velocity dispersion. We use the above 
scaling relations to combine appropriate subsamples of the HMF 
and maxBCG clusters and produce a conservative merged catalog of 799 clusters of galaxies 
at z$_{est}$ = 0.05 - 0.3 above richness threshold
of \lam $\geq$ 40 (HMF) and \ngal $\geq$ 13 (maxBCG) (\S \ref{catalog}). This threshold corresponds 
to clusters with a typical mean velocity dispersion of \sigv $\ga$ 400 km s$^{-1}$, total $r$-band 
luminosity L$^{{}^{tot}}_{0.6} \ga 3\times10^{11} h^{-2}$ \lsun\ and mass M$_{0.6} \ga 5\times10^{13} h^{-1}$ 
\msun\ (within a radius of 0.6 $h^{-1}$ Mpc). This threshold reflects clusters that are
poorer than Abell richness class 0. The average space density of the clusters
is $2\times10^{-5}\ h^3$  clusters/Mpc$^3$. Using the relevant selection functions, we determine 
the cluster richness function; we find it to be a steeply declining function of cluster abundance with increasing 
richness. We compare the cataloged clusters with the Abell and X-ray clusters 
located in the survey region; they are all detected (with 5 of the 58 Abell clusters below the 
above merged richness cuts).

The relevant selection functions for the catalog clusters are provided.
The catalog can be used for studies of individual clusters, for comparisons
with other objects (e.g., X-ray clusters, SZ clusters, AGNs), and in statistical
analyses (when properly corrected for the relevant selection functions). 

As an example, we determined the mass function of clusters (see \citealt{bahn03})
and used it to place powerful constarints on the mass-density parameter of the universe and the
amplitude of mass fluctuations; we find \om = 0.19 $\pm^{0.08}_{0.07}$ and $\sigma_8$ = 0.9 $\pm^{0.3}_{0.2}$.

The current work represents preliminary results from early SDSS commissioning data 
(4$\%$ of the ultimate SDSS survey). The results will greatly improve as more extensive 
SDSS data become available.




\acknowledgments

The SDSS is a joint project of The University of Chicago, Fermilab, the Institute for Advanced Study, the Japan 
Participation Group, The Johns Hopkins University, Los Alamos National Laboratory, the Max-Planck-Institute 
for Astronomy (MPIA), the Max-Planck-Institute for Astrophysics (MPA), New Mexico State University, 
University of Pittsburgh, Princeton University, the United States Naval Observatory, and the University of 
Washington.

Funding for the creation and distribution of the SDSS Archive has been provided by the Alfred P.
Sloan Foundation, the Participating Institutions, the National Aeronautics and Space 
Administration, the National Science Foundation, the U.S. Department of Energy, the Japanese 
Monbukagakusho, and the Max Planck Society. The SDSS Web site is http://www.sdss.org/. Tim McKay acknowledges 
support from NSF PECASE grant AST 9708232.




\clearpage

\epsscale{0.9}
\begin{figure}
\plotone{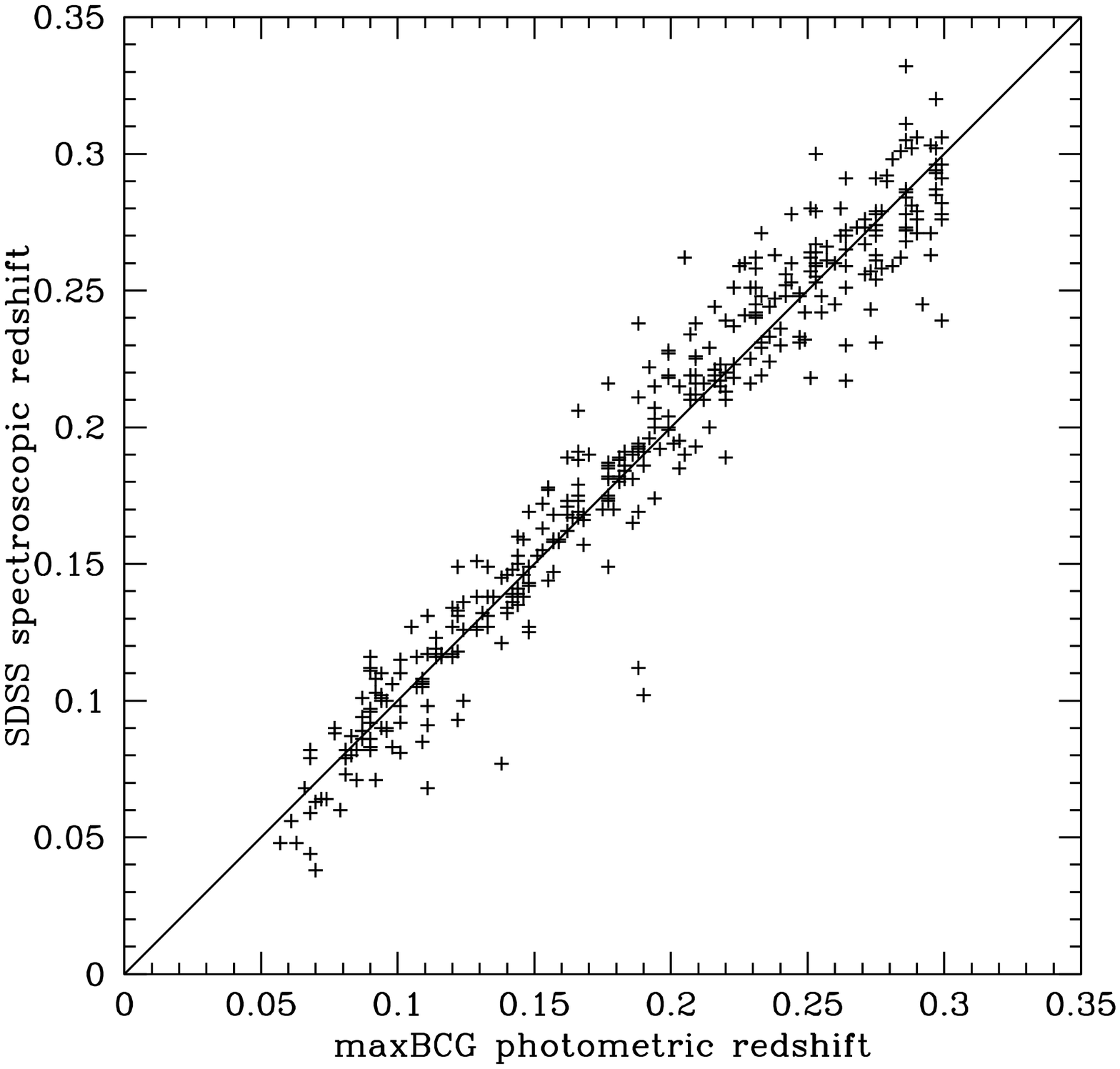}
\caption{Comparison of measured SDSS spectroscopic redshifts with  
photometric redshifts estimated by the maxBCG method for 382 
maxBCG clusters (\ngal $\geq$13, z$_{est}$ = 0.05-0.3). The dispersion 
in the estimated redshifts is $\sigma_z$ = 0.014. 
\label{f1}}
\end{figure}

\epsscale{0.9}
\begin{figure}
\plotone{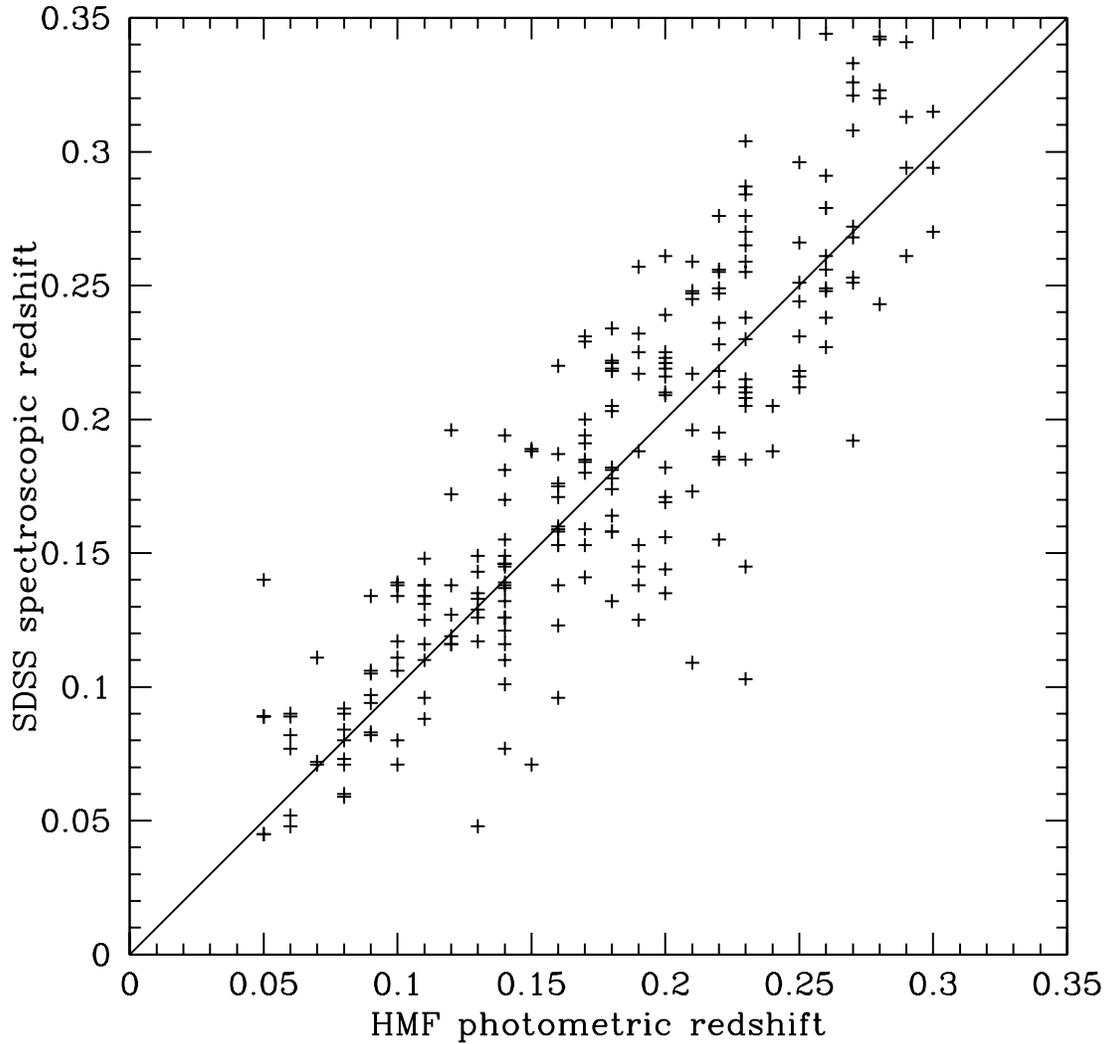}
\caption{Comparison of measured SDSS spectroscopic redshifts with
photometric redshifts estimated by the HMF method for 237 
HMF clusters (\lam $\geq$40, z$_{est}$ = 0.05-0.3). The dispersion 
in the estimated redshifts is $\sigma_z$ = 0.033. 
\label{f2}}
\end{figure}

\epsscale{0.9}
\begin{figure}
\plotone{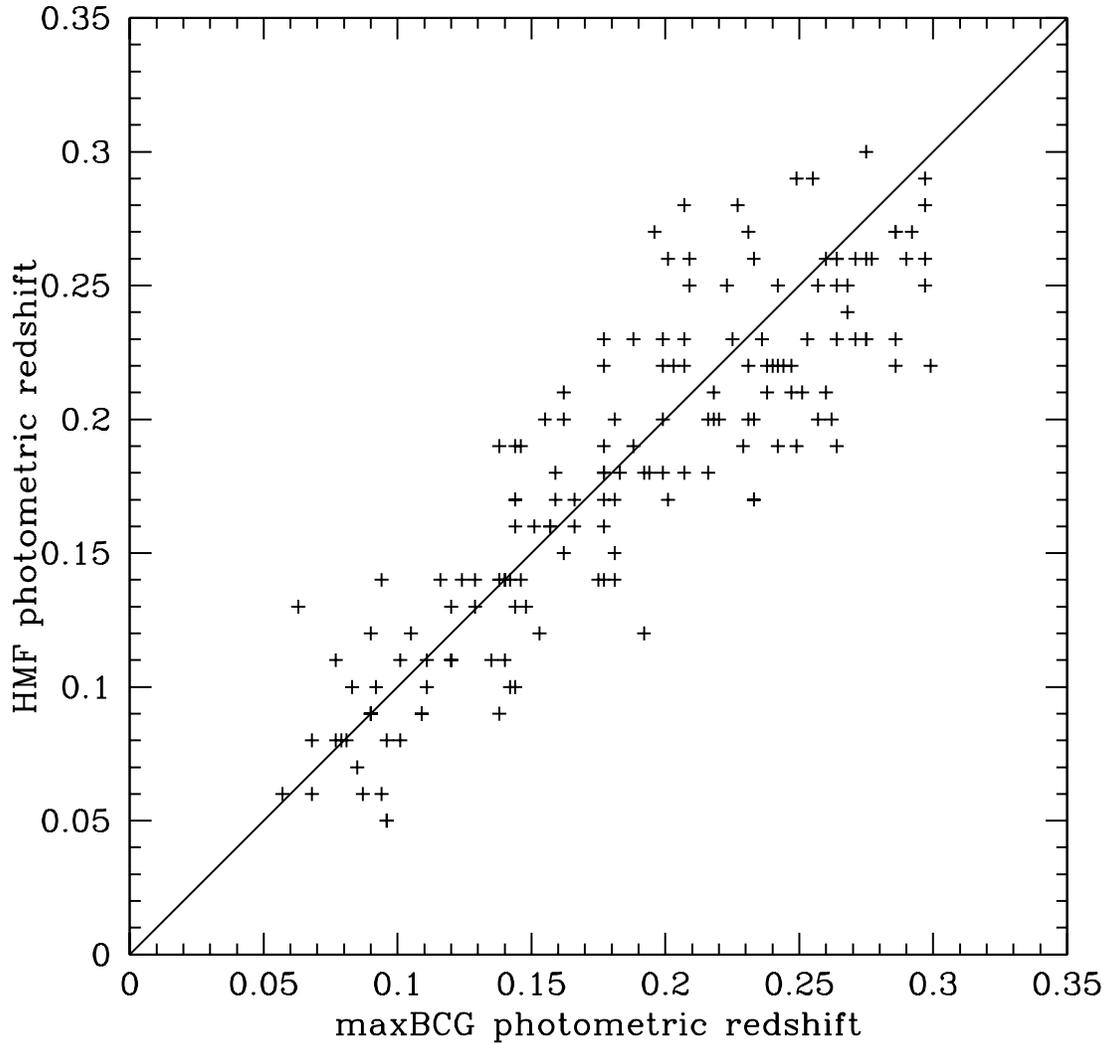}
\caption{Comparison of HMF and maxBCG estimated redshifts for 161 
cluster pairs (\lam $\geq$40, \ngal $\geq$13, z$_{est}$ = 0.05-0.3). The 
cluster pairs are separated by $\leq 1 h^{-1}$ Mpc (projected) and $\Delta$z$_{est}$ $\leq$ 0.08 in estimated 
redshift. 
\label{f3}}
\end{figure}

\epsscale{0.8}
\begin{figure}
\plotone{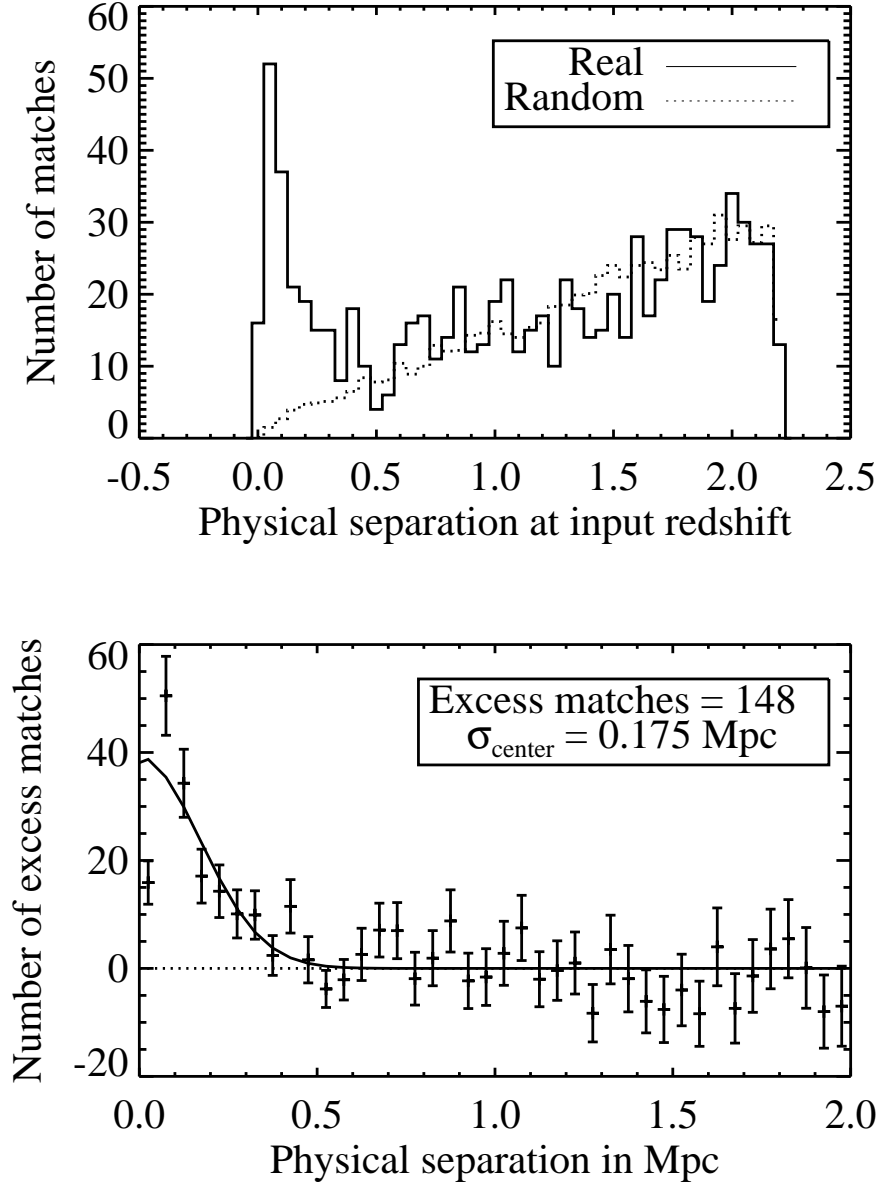}
\caption{Top panel is the histogram of the number of matches between 
HMF and maxBCG clusters (z = 0.05 - 0.3) as a function of physical projected separation 
in Mpc (calculated at the maxBCG estimated redshift). The solid line represents 
the data (real matches); the dashed line results from matching the clusters 
with random positions (thus representing chance contribution to matches).
Lower panel shows the difference between these two (all matches minus
random matches). The excess pairs are concentrated at small separations 
($\la$ 0.5 $h^{-1}$ Mpc) and represent real matches. 
\label{f4}}
\end{figure}

\epsscale{0.9}
\begin{figure}
\plotone{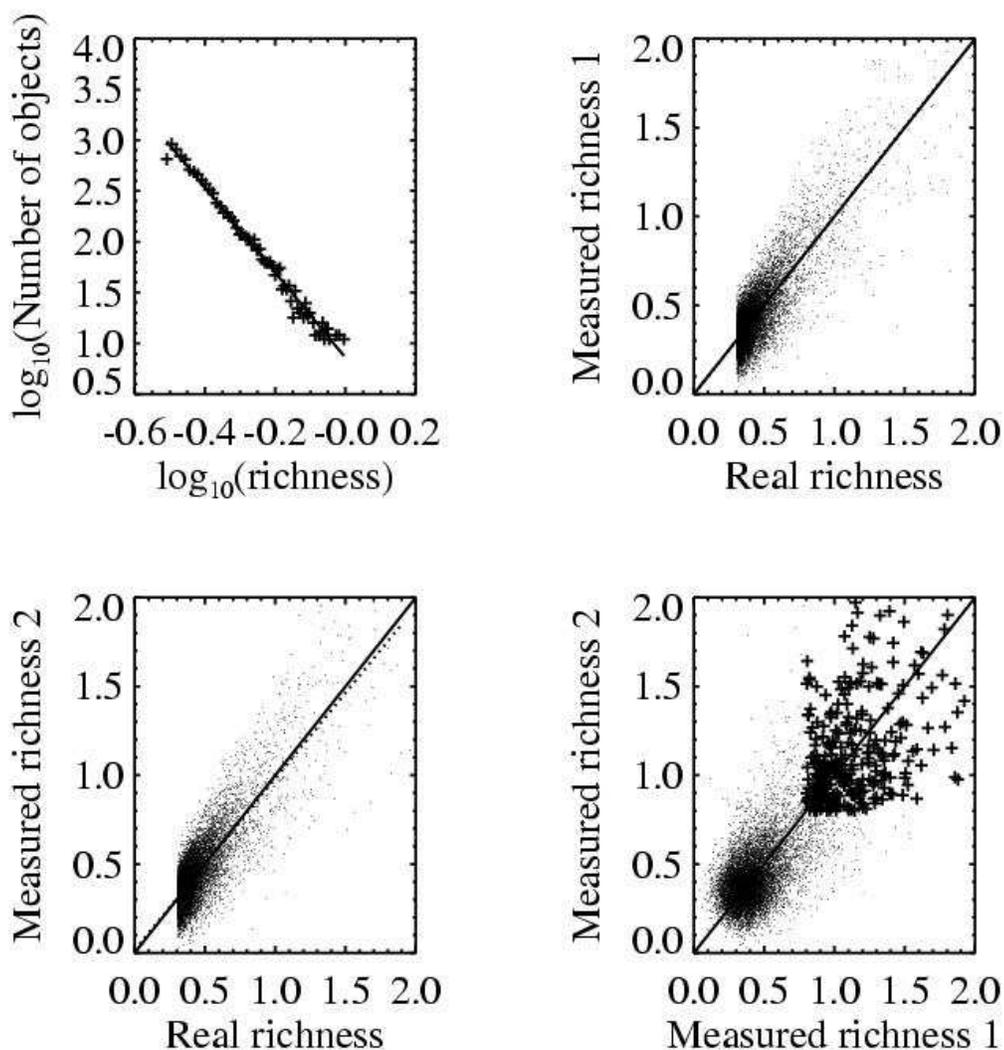}
\caption{Monte Carlo simulations showing the effect of uncertainty in richness 
estimates on comparison of catalogs drawn from a steeply declining 
richness function. The top left panel shows the model richness function
(N$_{cl} \sim$ Richness$^{-4}$). The top right and bottom left panels compare measured
to actual richness measures for two realizations of richness measurements 
with 30$\%$ measurement uncertainties. The bottom right panel compares the 
richness measurements of the two Monte Carlo realizations of the data, 
illustrating that only 54$\%$ of the clusters passing one richness threshold 
will also pass the other. 
\label{f5}}
\end{figure}

\epsscale{0.7}
\begin{figure}
\plotone{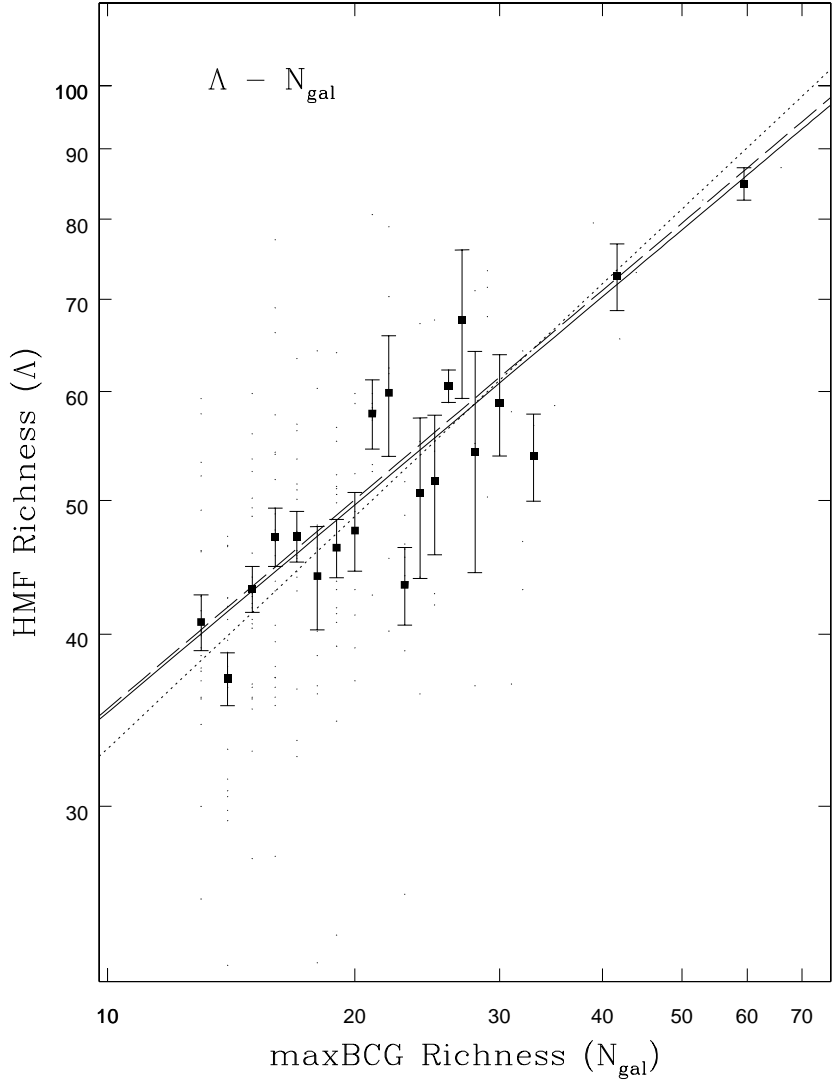}
\caption{Comparison of HMF and maxBCG richnesses. The HMF richness
\lam\ (determined for HMF clusters) is compared with the maxBCG richness
\ngal\ (determined for maxBCG clusters with \ngal$\geq$13) for matched cluster pairs 
(HMF clusters that match maxBCG clusters within 1 $h^{-1}$ Mpc projected separation and $\Delta$z $\leq$ 0.05). 
Individual \lam-\ngal\ matches are shown by the faint points; 
the mean richness \lam\ as a function of \ngal\ is presented by the solid squares, with rms error-bars on the means. 
The best-fit relation, 
\lam = (11.1$\pm$0.8) \ngal$^{0.5\pm0.03}$, is shown by the solid line.
The dashed line represents the independent correlation obtained using 
the observed luminosity-richness relations for HMF and maxBCG clusters
(\S \ref{scaling}, figures 7 and 8). The dotted line represents another independent relation 
implied from the observed velocity dispersion versus richness correlations 
(\S \ref{scaling}, figure 9). All three independent methods yield consistent results.
\label{f6}}
\end{figure}

\epsscale{0.7}
\begin{figure}
\plotone{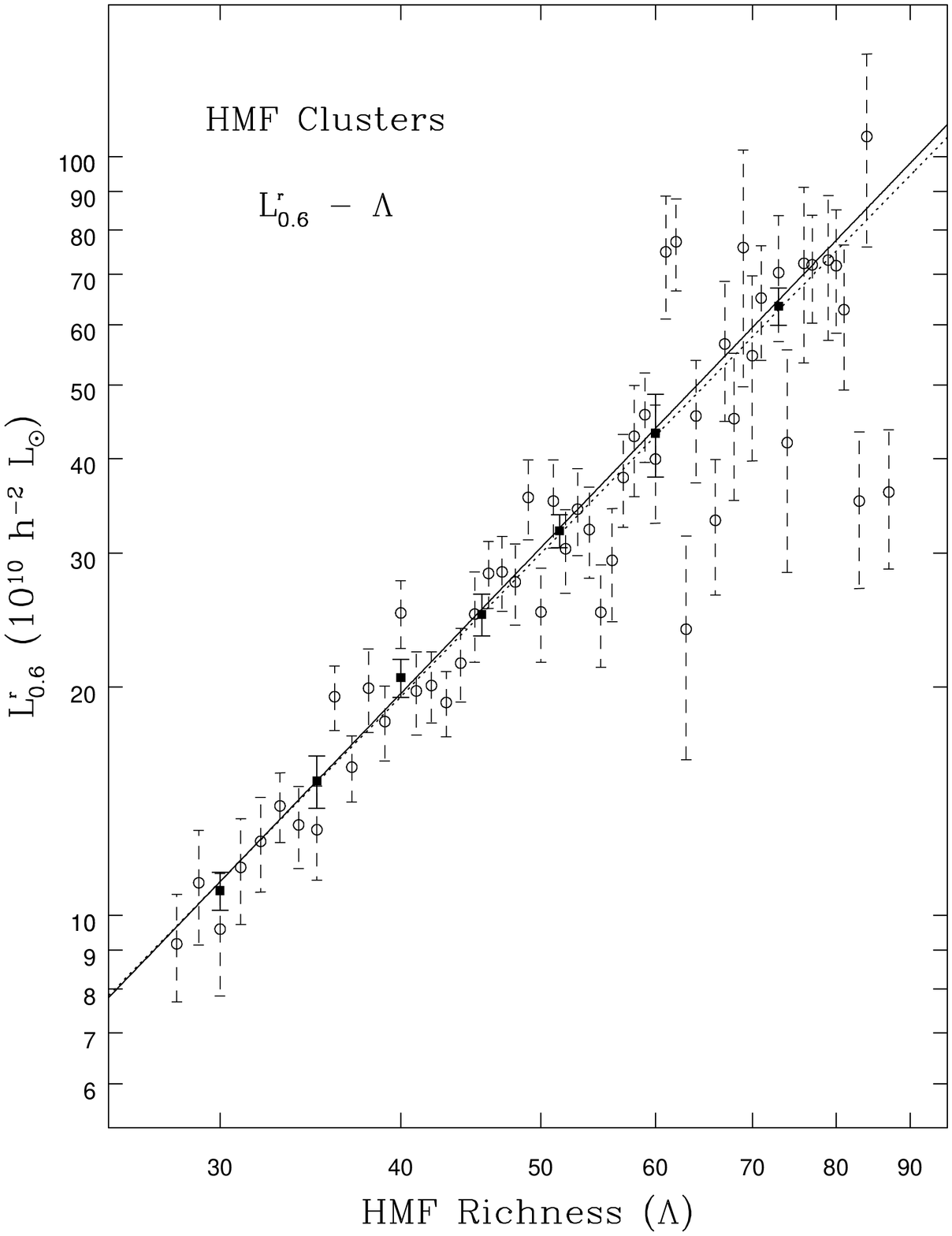}
\caption{Observed cluster luminosity versus richness for HMF clusters.
Cluster luminosity is observed in the $r$-band, within a radius of 0.6 $h^{-1}$ Mpc,
for stacked clusters at a given richness. The luminosities are k-corrected,
background subtracted, and integrated down to M$_r$ = $-19.8$. Dark squares 
represent binned data (in richness bins) of the stacked clusters. The solid line is the best-fit power-law relation
(for the range \lam $\simeq$ 30 - 80): $L^{{}^r}_{0.6} (10^{10} L_\sun) = 0.013\ \Lambda^{1.98}$ 
(Equation 3). (The dotted line is the best-fit when the \lam$>$80 higher scatter
clusters are added). The contribution of galaxies
fainter than $-19.8$ adds a correction factor of 1.42 to the above luminosities
(\S \ref{scaling}). 
\label{f7}}
\end{figure}

\epsscale{0.7}
\begin{figure}
\plotone{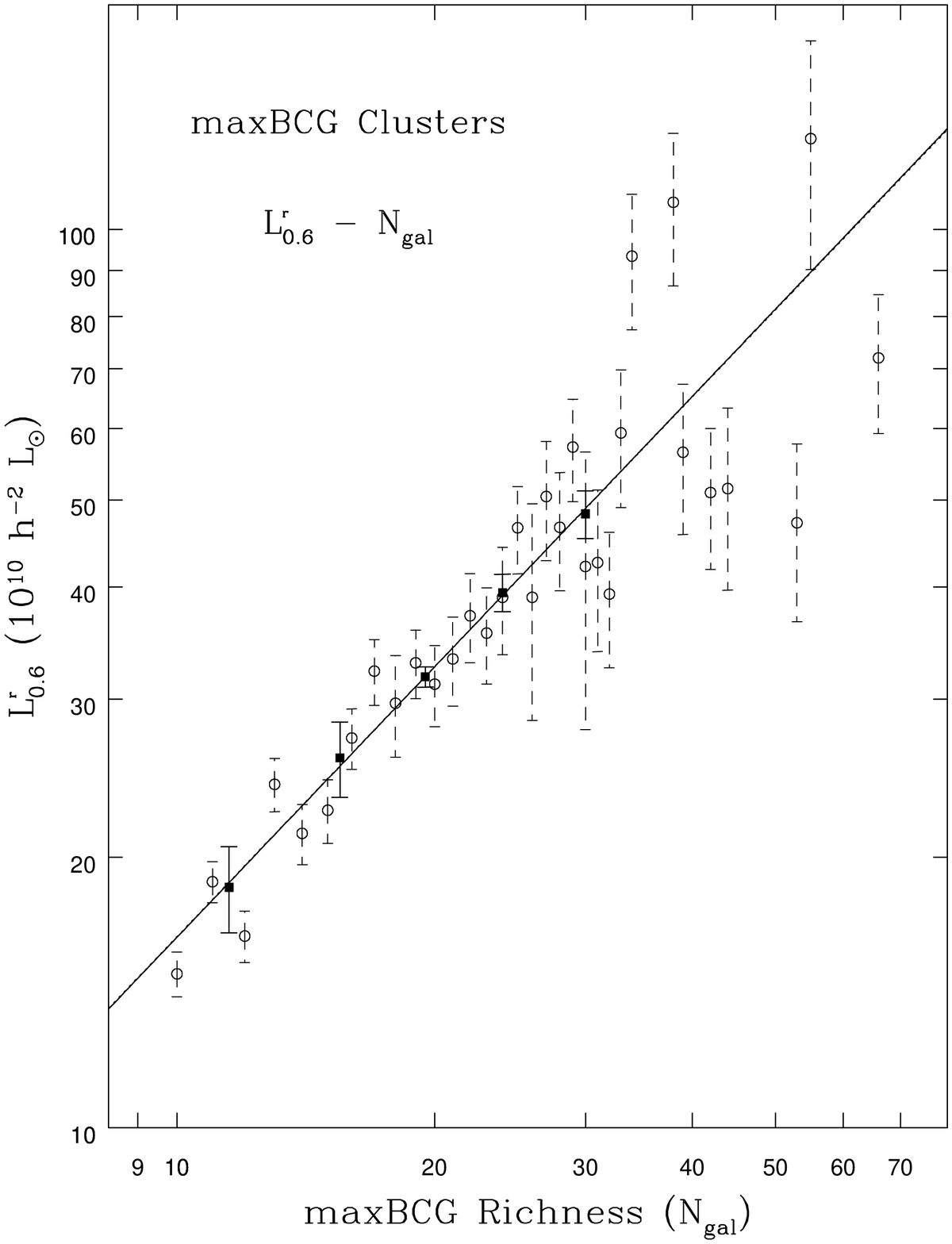}
\caption{Observed cluster luminosity versus richness for maxBCG clusters.
Cluster luminosity is observed in the $r$-band, within a radius of 0.6 $h^{-1}$ Mpc,
for stacked clusters at a given richness. The luminosities are k-corrected,
background subtracted, and integrated down to M$_r$ = $-19.8$. Dark squares 
represent binned data (in richness bins) of the stacked clusters. The solid line is the best-fit power-law relation
(for the range \ngal $\simeq$ 10 - 33): $L^{{}^r}_{0.6} (10^{10} L_\sun) = 1.6\ N_{gal}$ 
(Equation 2). (A similar relation is obtained when the \ngal$>$33 higher scatter
clusters are added, shown by the dotted line which overlaps the solid line). The contribution of galaxies
fainter than $-19.8$ adds a correction factor of 1.34 to the above luminosities
(\S \ref{scaling}).
\label{f8}}
\end{figure}

\epsscale{0.73}
\begin{figure}
\plotone{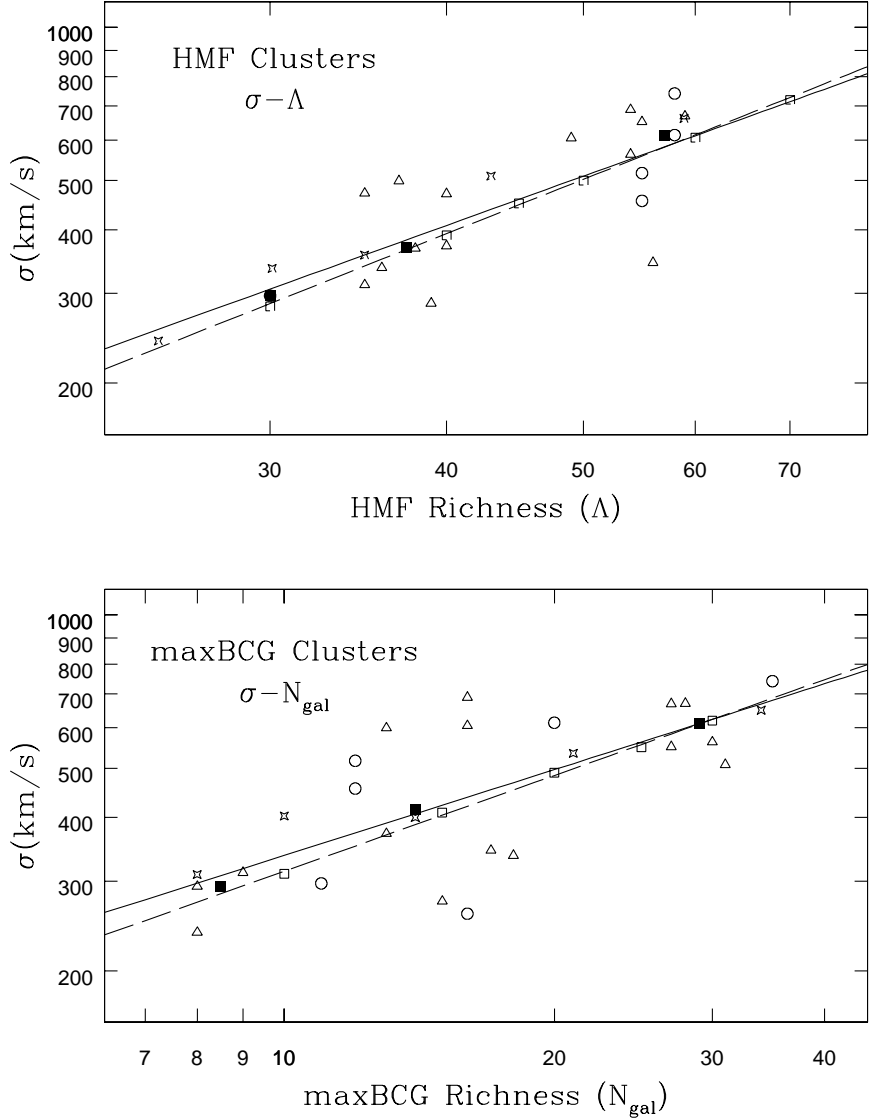}
\caption{Relation between observed cluster velocity dispersion $\sigma$ and
cluster richness. Triangles are SDSS observed velocity dispersions, 
circles are Abell clusters, dark squares are medians, and the solid line
is the best fit to the velocity data. Stars represent SDSS observations
of Gaussian $\sigma$ from stacked galaxy velocity differences (relative to the BCG velocity) in all clusters 
with available data (shown for comparison only). Typical uncertainties in the 
velocity dispersion measurements and the richness estimates are $\sim$20$\%$ (1-$\sigma$). 
The dashed line represents the expected relation based on the observed luminosity-richness relations
(Figures 7 and 8) [followed by a conversion of luminosity to mass using 
mean $M/L$ ratios and a conversion of mass to velocity dispersion using 
observed gravitational lensing calibration; see \citealt{bahn03}].
\label{f9}}
\end{figure}

\epsscale{1.0}
\begin{figure}
\plotone{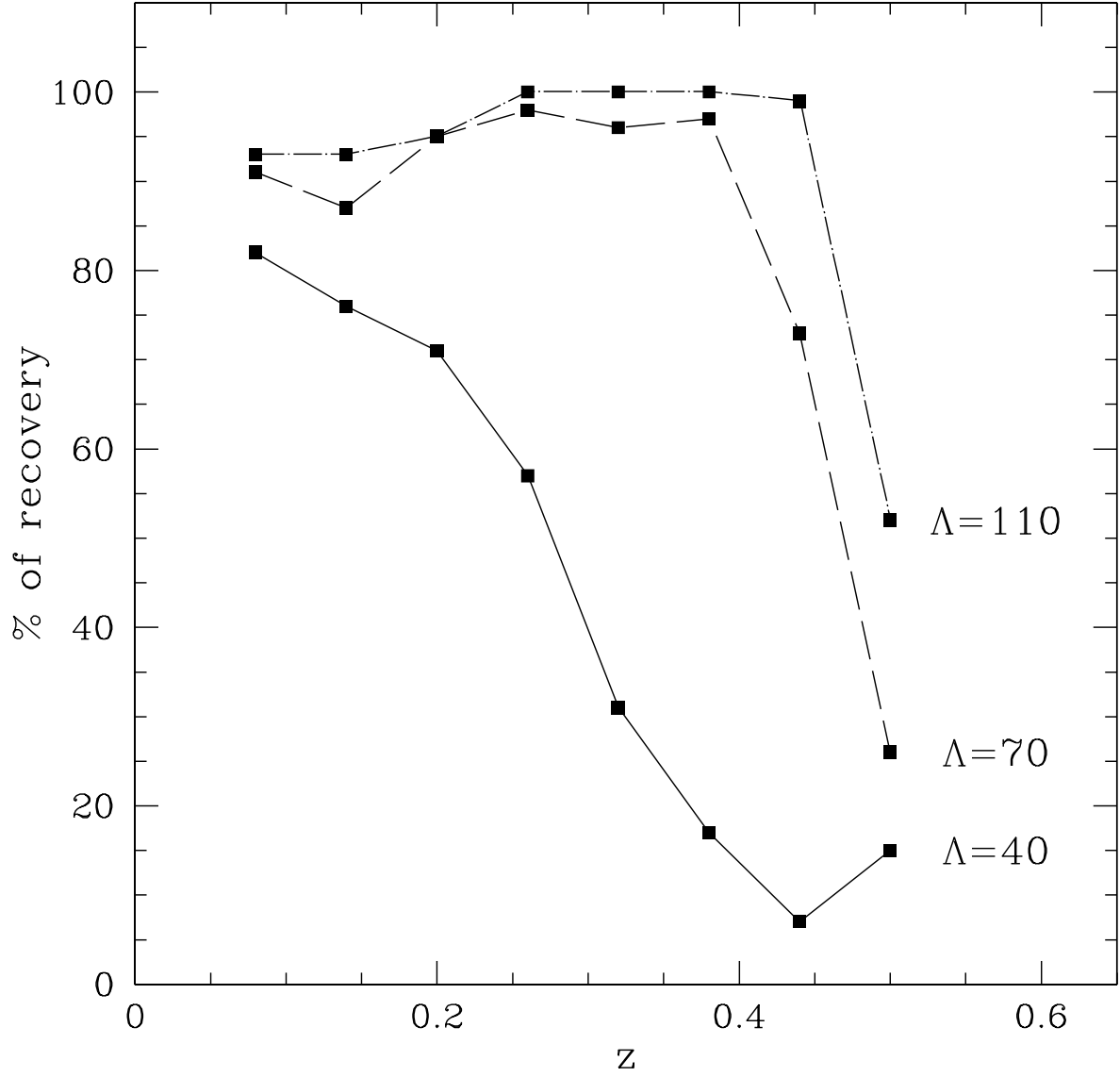}
\caption{Selection function for HMF clusters as a function of redshift 
and richness; determined from cluster simulations \citep{kim02a}.
\label{f10}}
\end{figure}

\epsscale{1.0}
\begin{figure}
\plotone{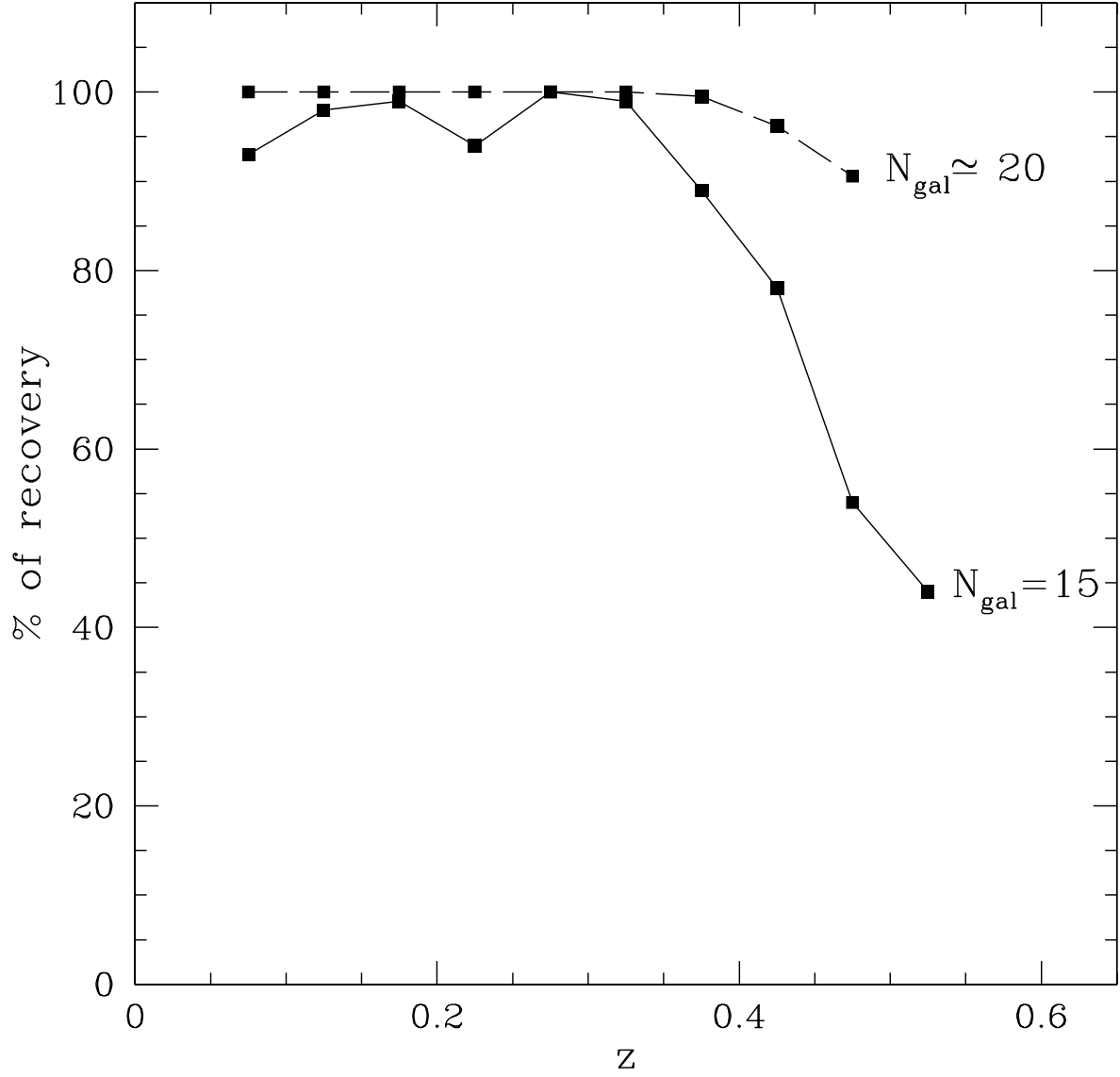}
\caption{Selection function for maxBCG clusters, determined from simulations
\citep{ann02}.
\label{f11}}
\end{figure}

\clearpage
\begin{figure}
\includegraphics[width=1.0\hsize,angle=90]{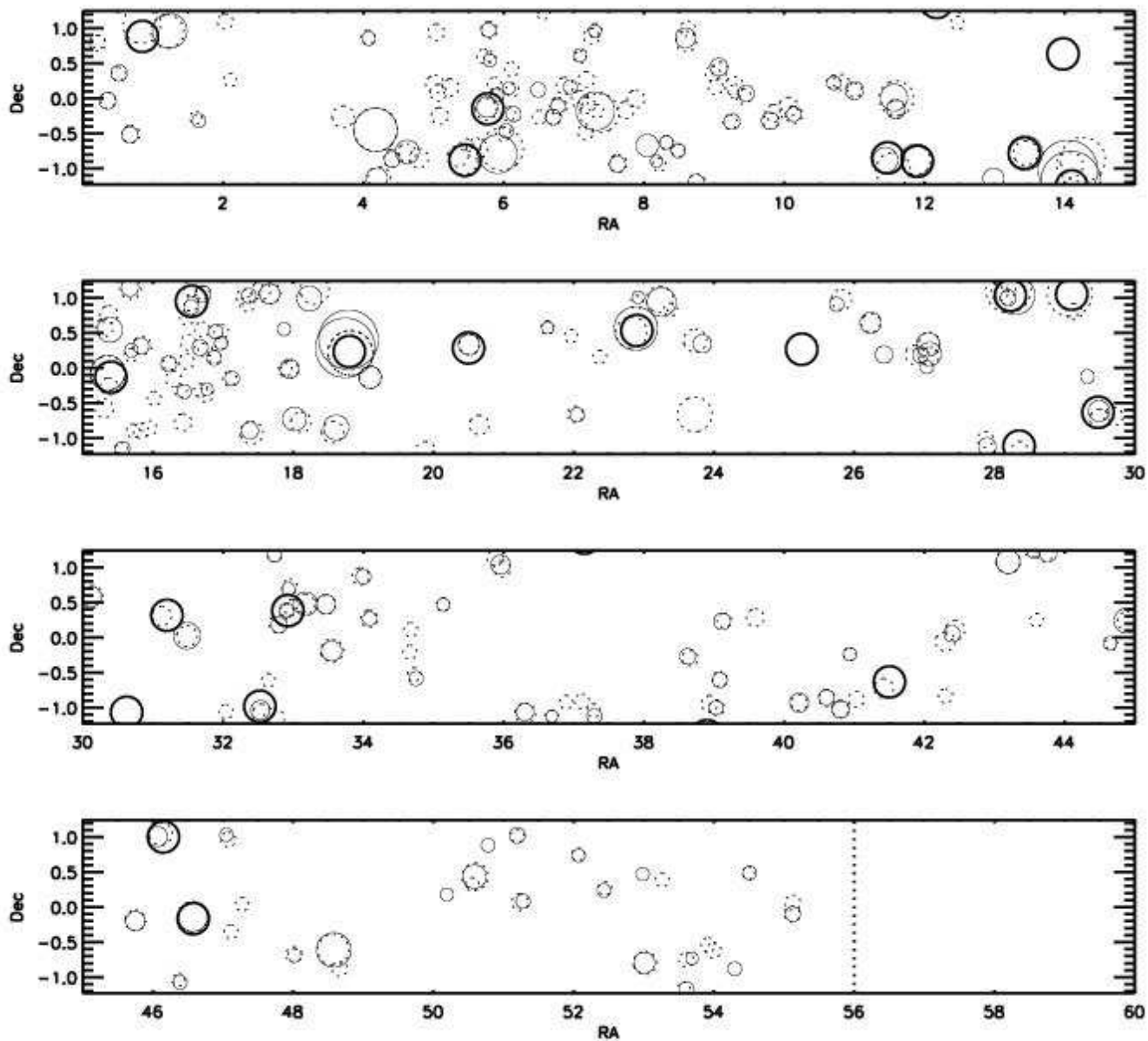}
\caption{Overlay map of clusters in the merged catalog
[z$_{est}$ = 0.05 - 0.3, \lam$\geq$40 (HMF), \ngal$\geq$13 (maxBCG), and their matches].
Dotted circles are maxBCG clusters, solid circles are HMF clusters, and bold solid 
circles are Abell clusters in the survey area. All circles have a radius
of 1 $h^{-1}$ Mpc at the cluster redshift. (A redshift of 0.1 is assumed for the Abell clusters, 
many of which have no measured redshift.)
\label{f12}}
\end{figure}
\begin{figure}
\includegraphics[width=1.0\hsize,angle=90]{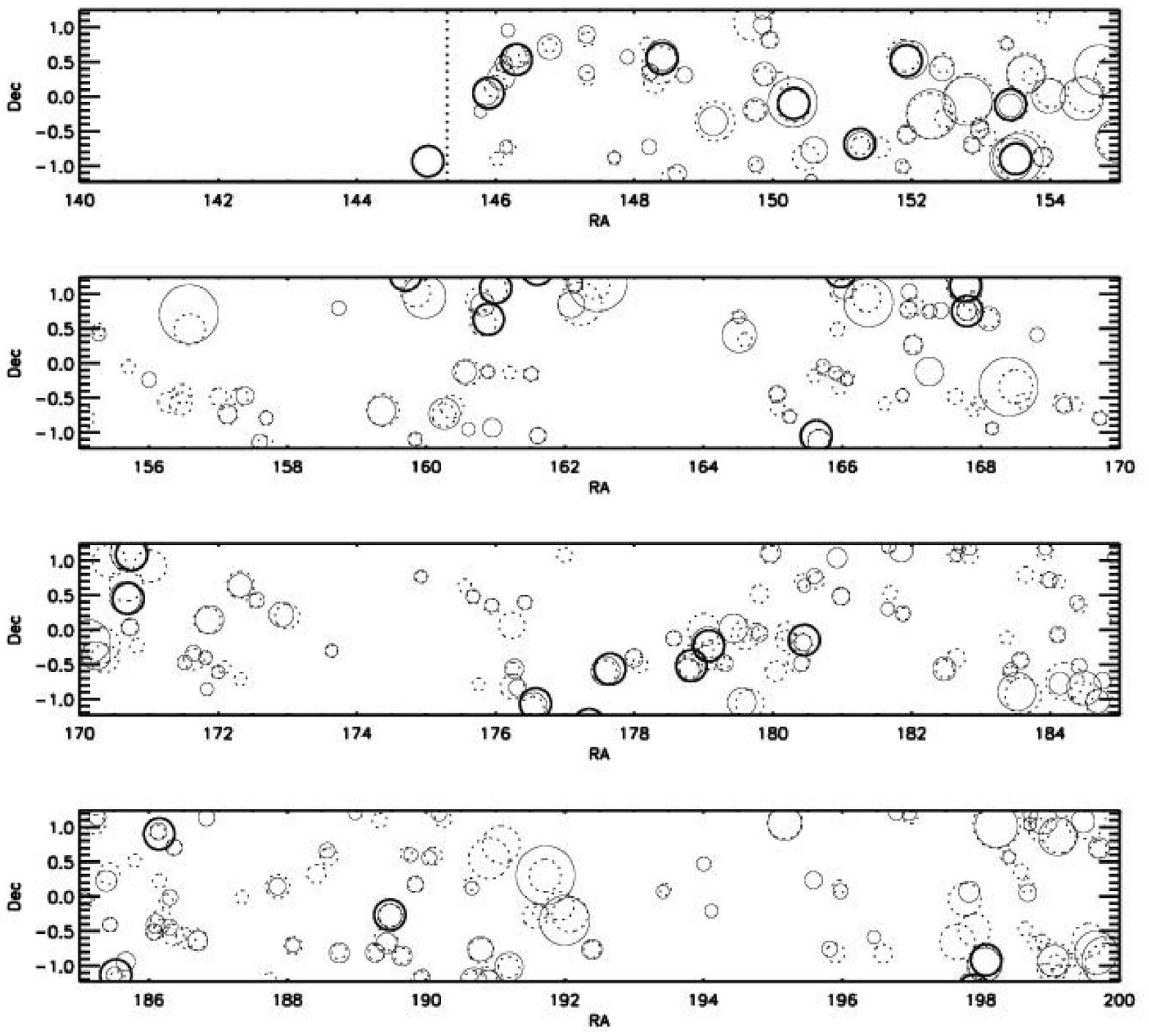}
\addtocounter{figure}{-1}
\caption{Continued}
\end{figure}
\begin{figure}
\includegraphics[width=1.0\hsize,angle=90]{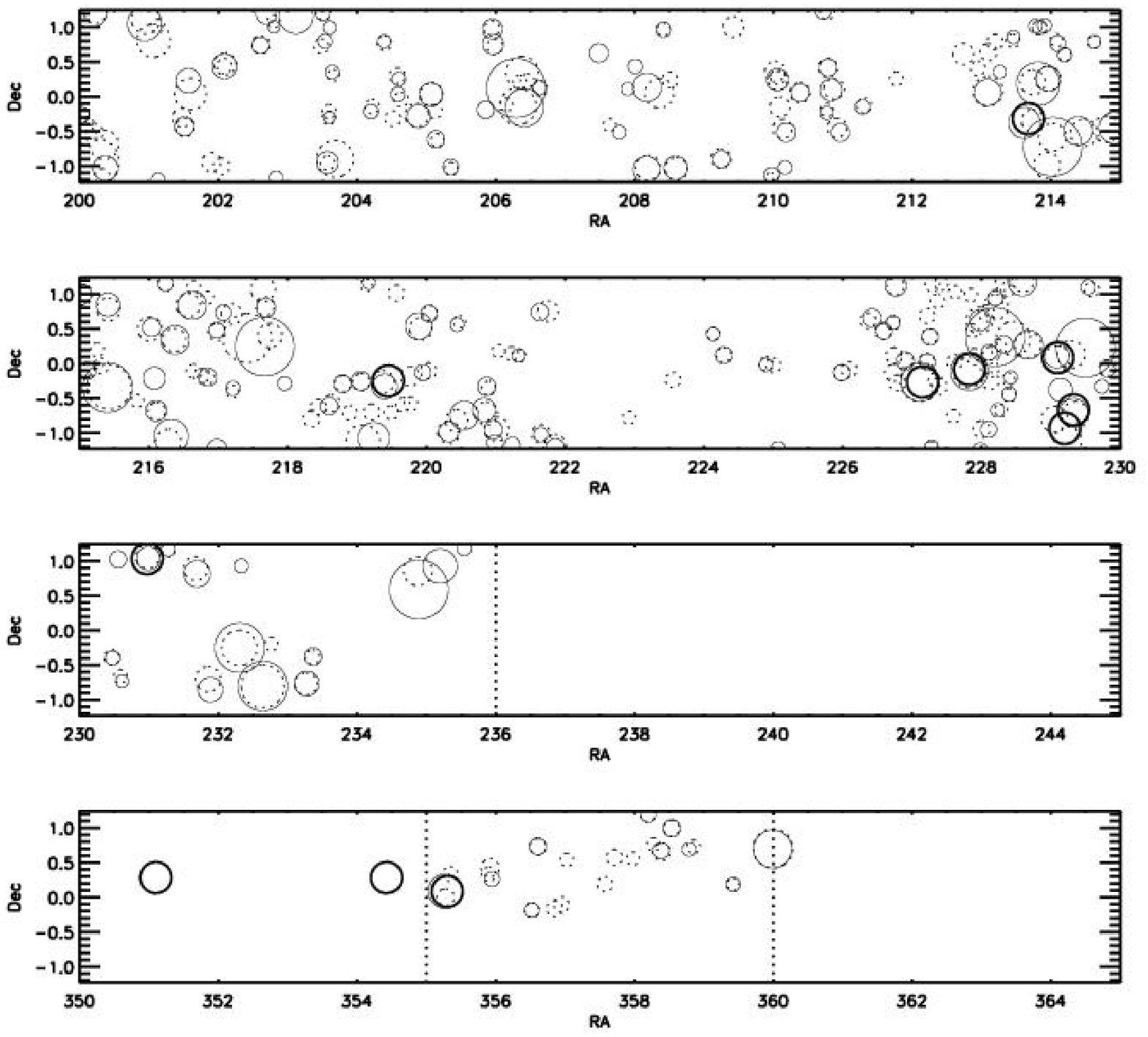}
\addtocounter{figure}{-1}
\caption{Continued}
\end{figure}

\begin{figure*}
\includegraphics[width=3.1in]{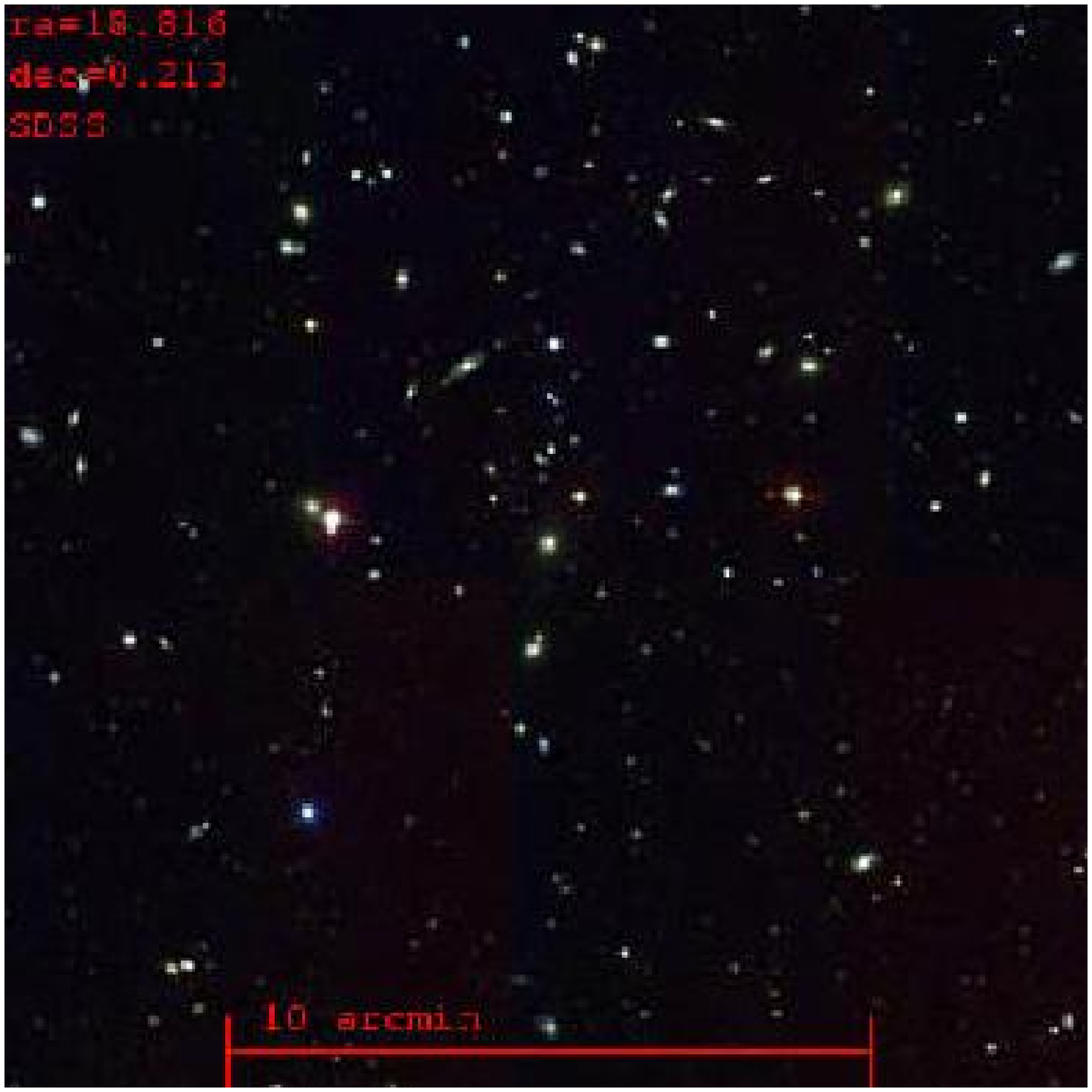}
\hspace{0.3in}
\includegraphics[width=3.1in]{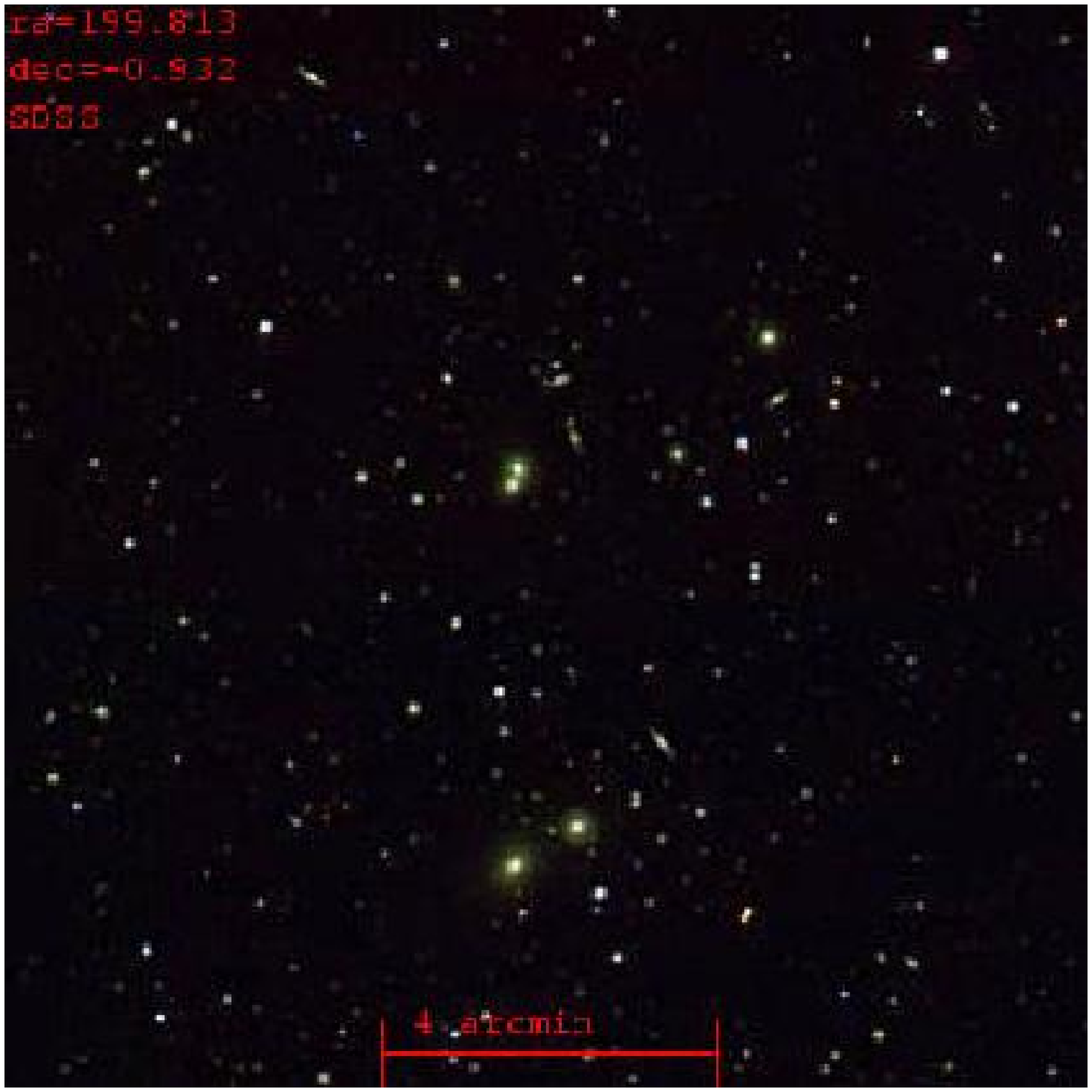}
\\
{\scriptsize BH148 (z=0.045; H,b; \lam=55.1; \ngal=12; A168; RXC0114.9)}  
\hskip 0.31in 
{\scriptsize BH563 (z=0.082; H,B; \lam=59.3; \ngal=27)}
\vskip 0.2in
\includegraphics[width=3.1in]{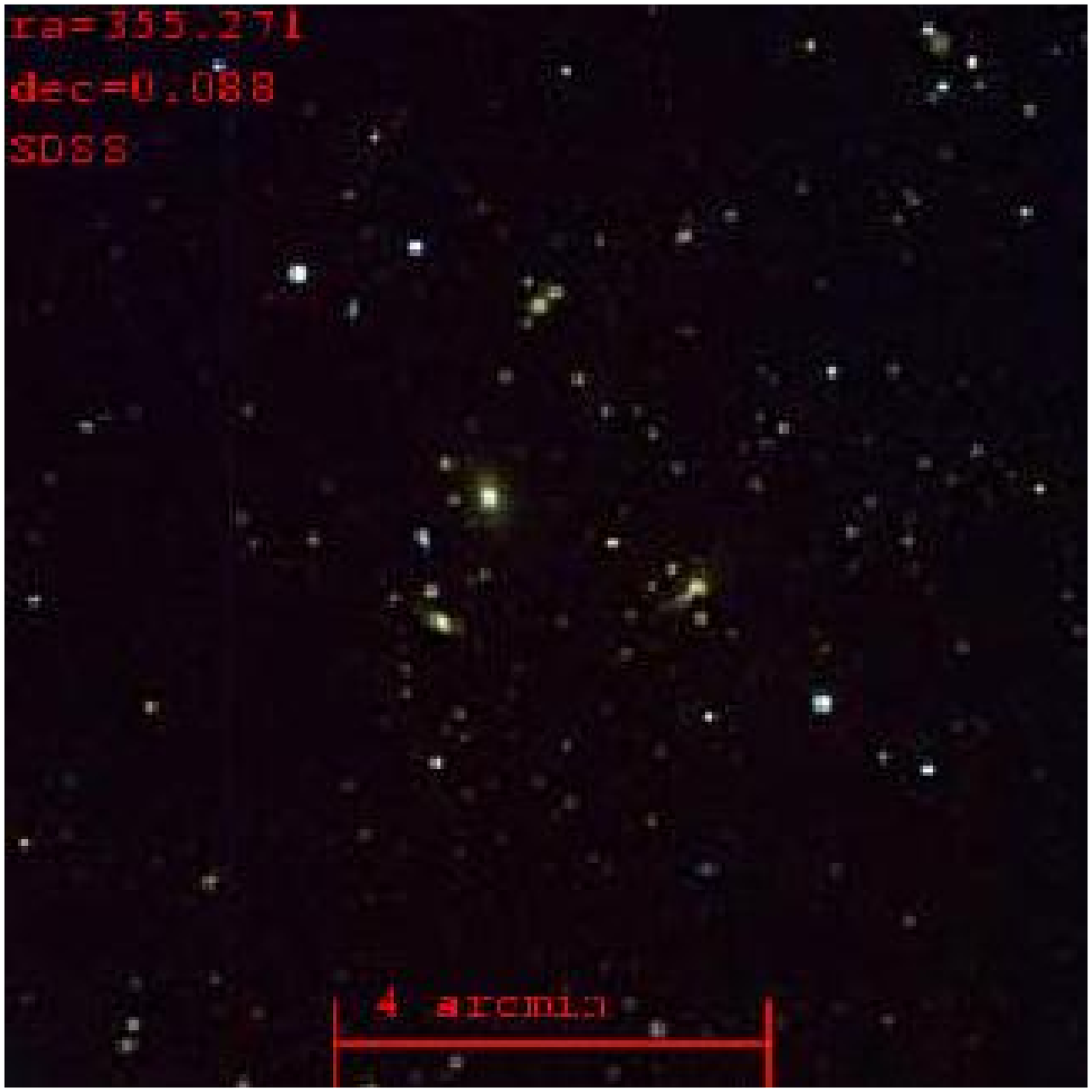}
\hspace{0.3in}
\includegraphics[width=3.1in]{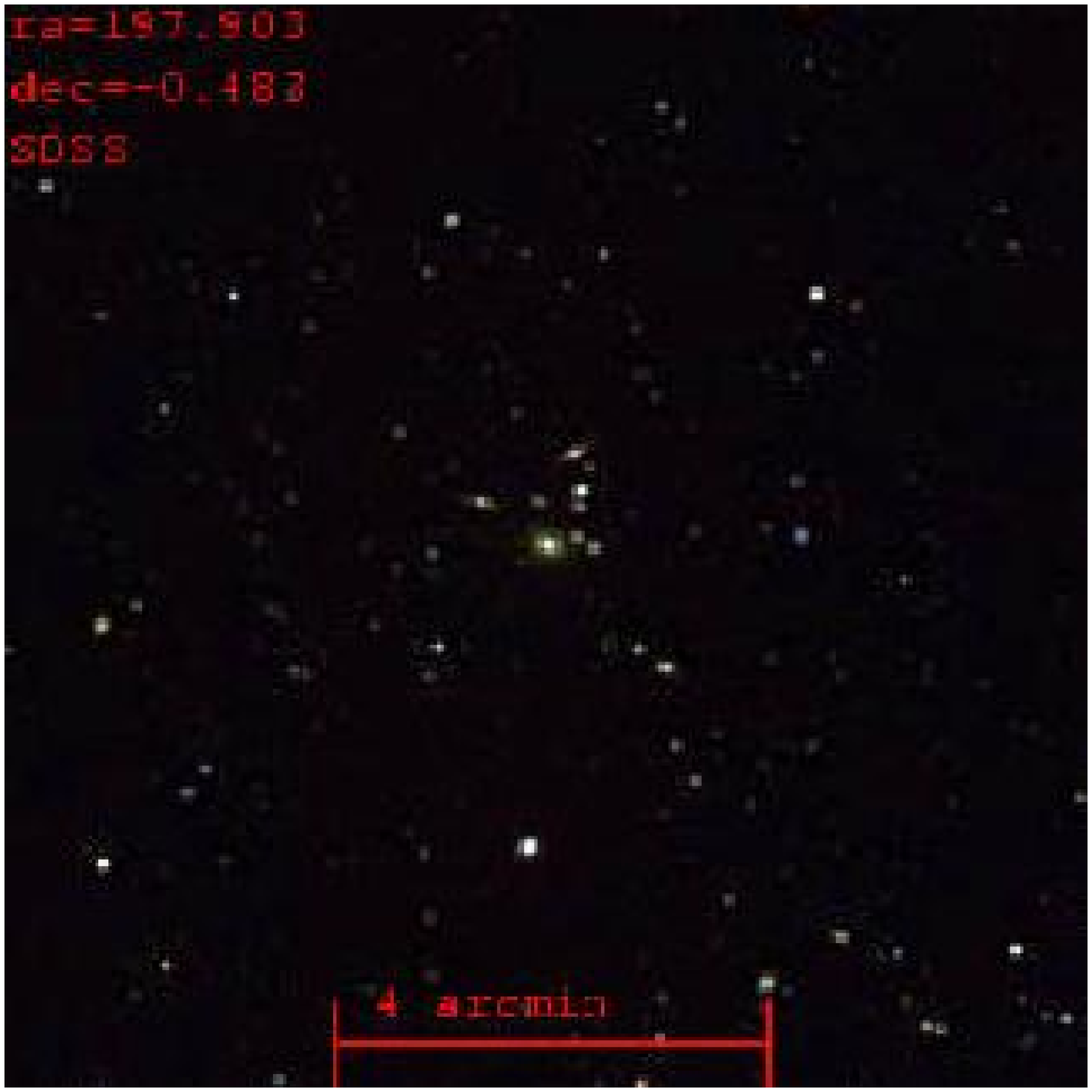}
\\
{\scriptsize BH1 (z=0.090; H; \lam=45.1; A2644; [RXC2341.1])} 
\hskip 0.92in 
{\scriptsize BH541 (z=0.090; B; \ngal=13)}
\vskip 0.2in
\caption{Images of a sample of cataloged clusters in the redshift range z $\simeq$ 0.05 - 0.3. 
The clusters are ordered by increasing redshift.
\label{f13}}
\end{figure*}
\begin{figure*}
\includegraphics[width=3.1in]{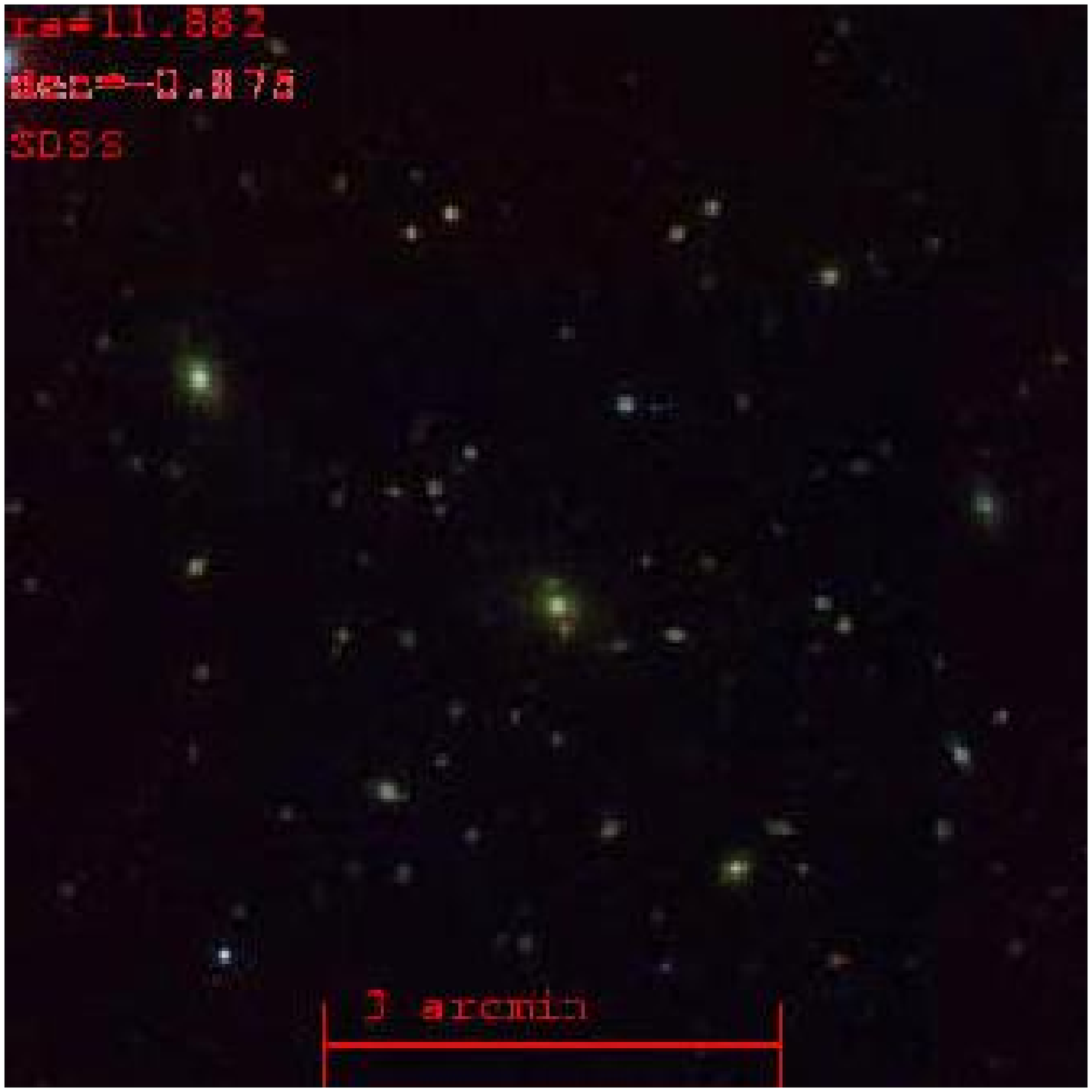}
\hspace{0.3in}
\includegraphics[width=3.1in]{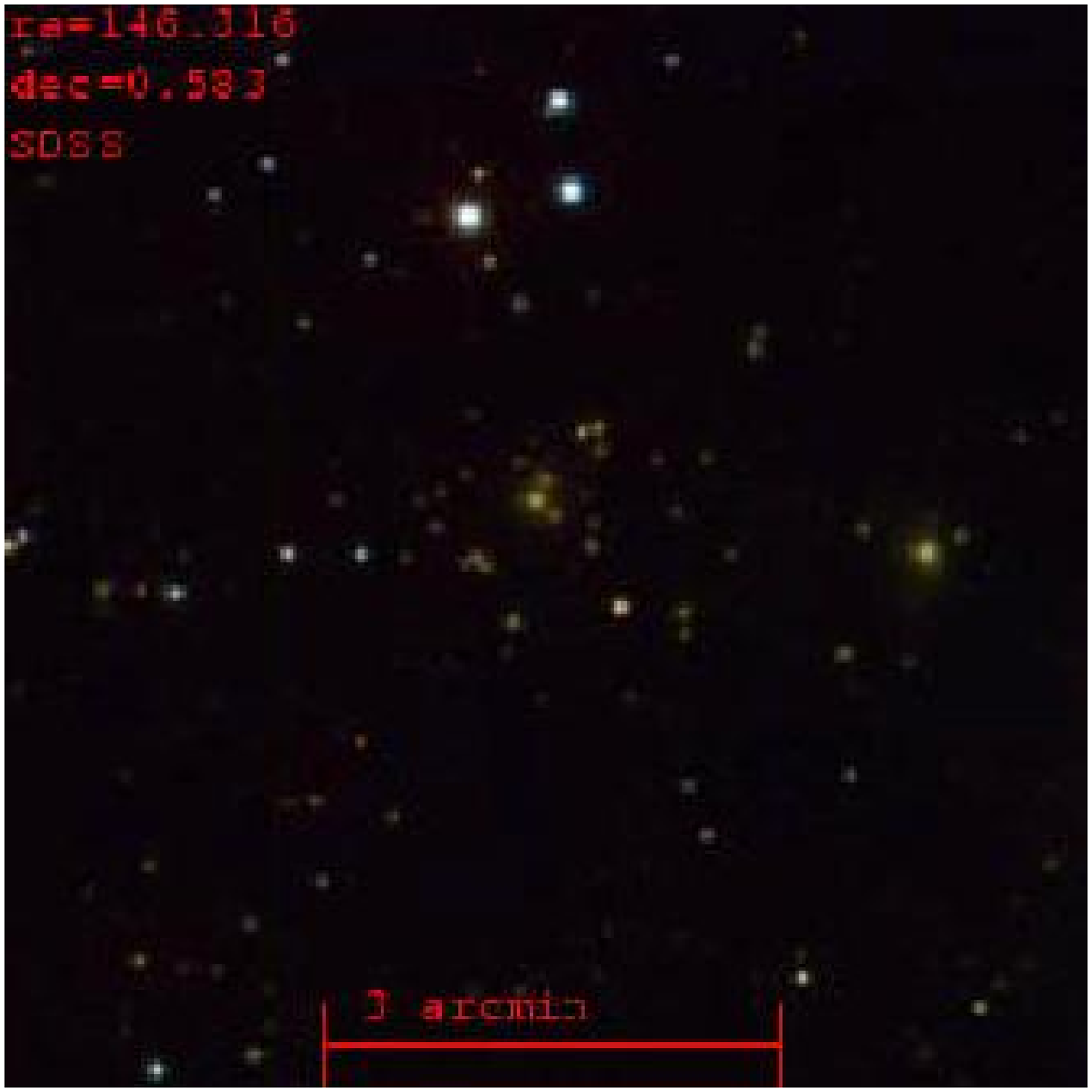}
\\
{\scriptsize BH100 (z=0.117; H,B; \lam=43.0; \ngal=17; A101)}  
\hskip 0.94in 
{\scriptsize BH269 (z=0.121; H,B; \lam=55.2; \ngal=19; A867)}
\vskip 0.2in
\includegraphics[width=3.1in]{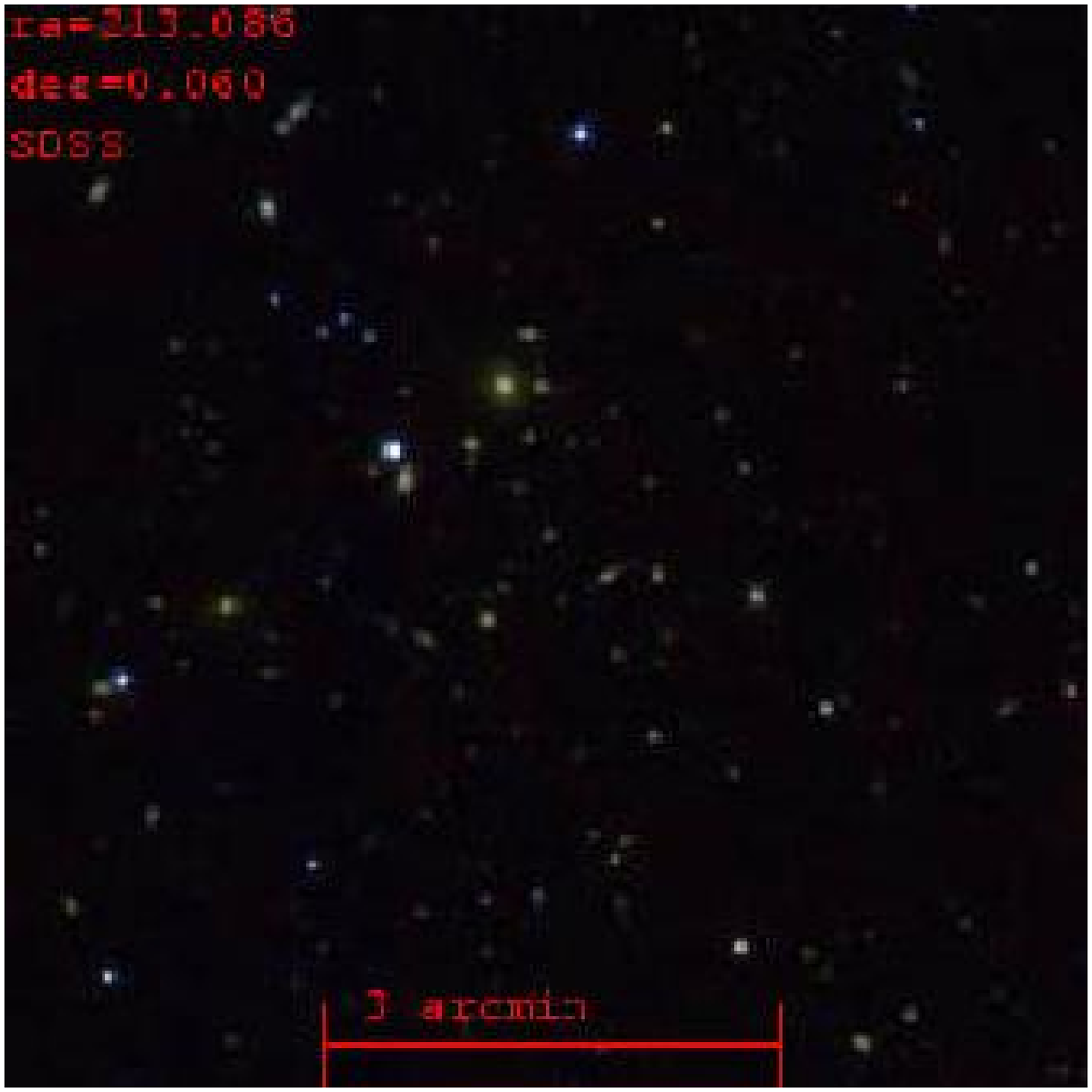}
\hspace{0.3in}
\includegraphics[width=3.1in]{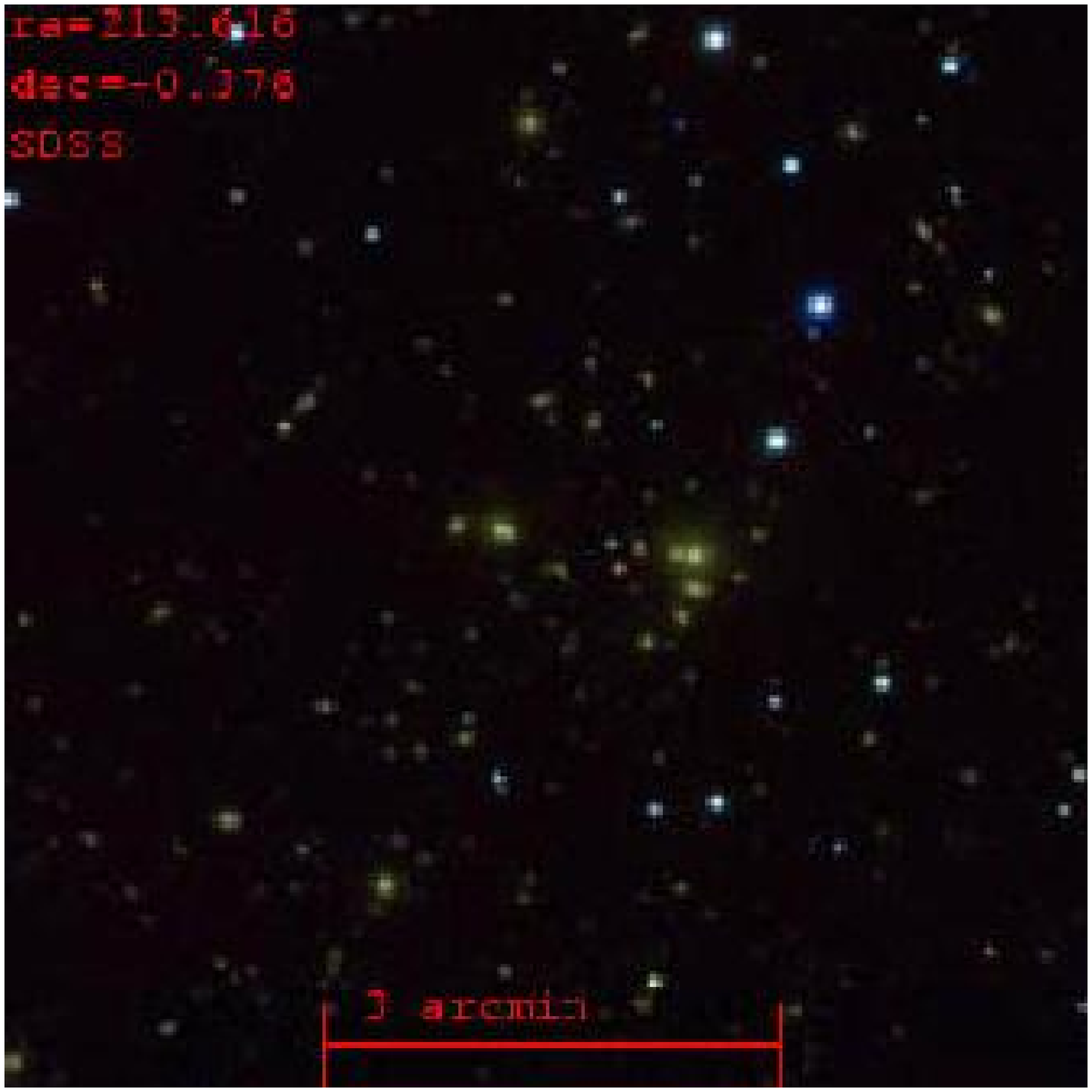}
\\
{\scriptsize BH645 (z=0.127; H,B; \lam=52.5; \ngal=15)}  
\hskip 1.28in 
{\scriptsize BH650 (z=0.139; H,B; \lam=87.2; \ngal=66; A1882)}
\vskip 0.2in
\addtocounter{figure}{-1}
\caption{Continued}
\end{figure*}
\begin{figure*}
\includegraphics[width=3.1in]{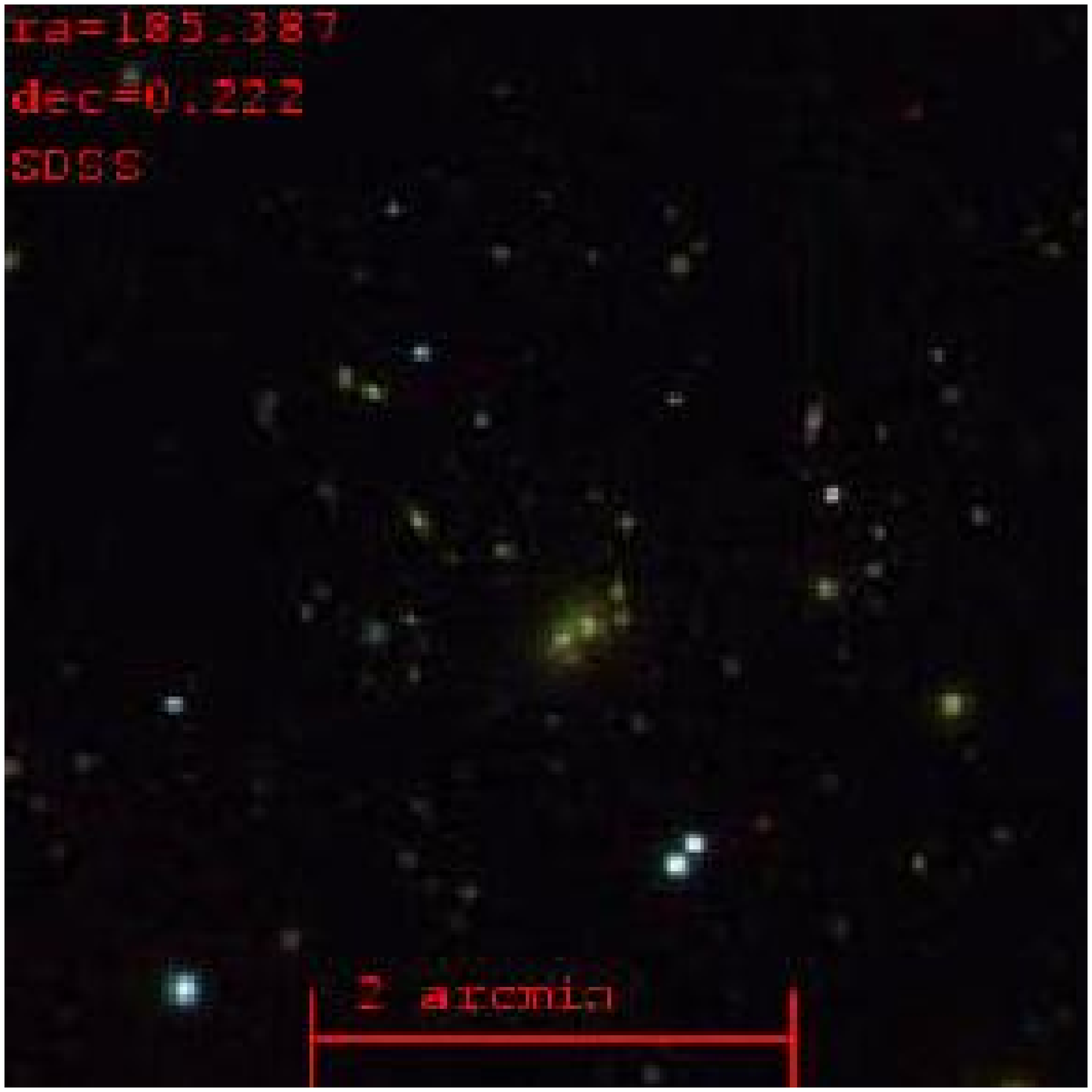}
\hspace{0.3in}
\includegraphics[width=3.1in]{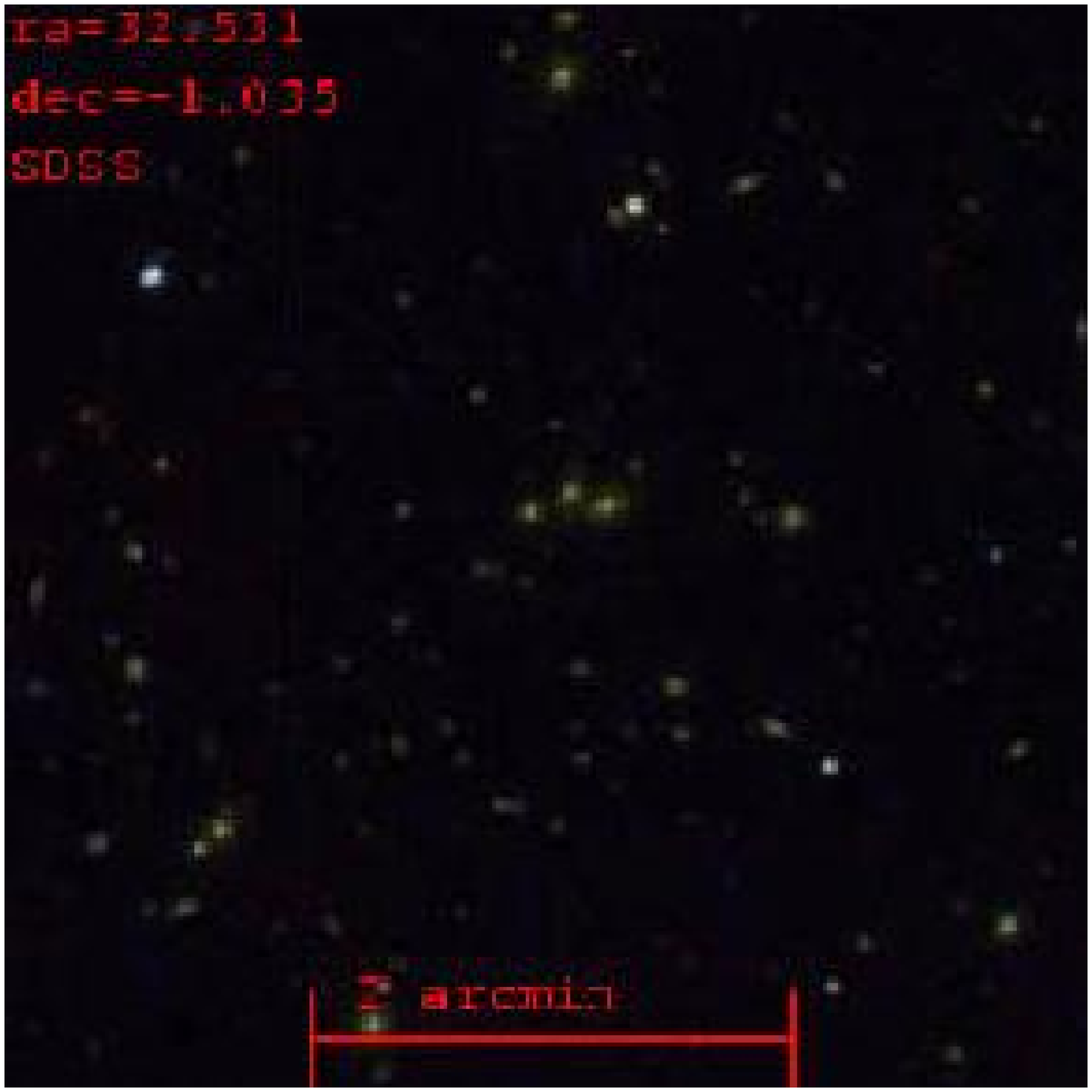}
\\
{\scriptsize BH479 (z=0.157; H,B; \lam=67.6; \ngal=25)}  
\hskip 1.28in 
{\scriptsize BH187 (z=0.174; H,B; \lam=67.2; \ngal=24; A315)}
\vskip 0.2in
\includegraphics[width=3.1in]{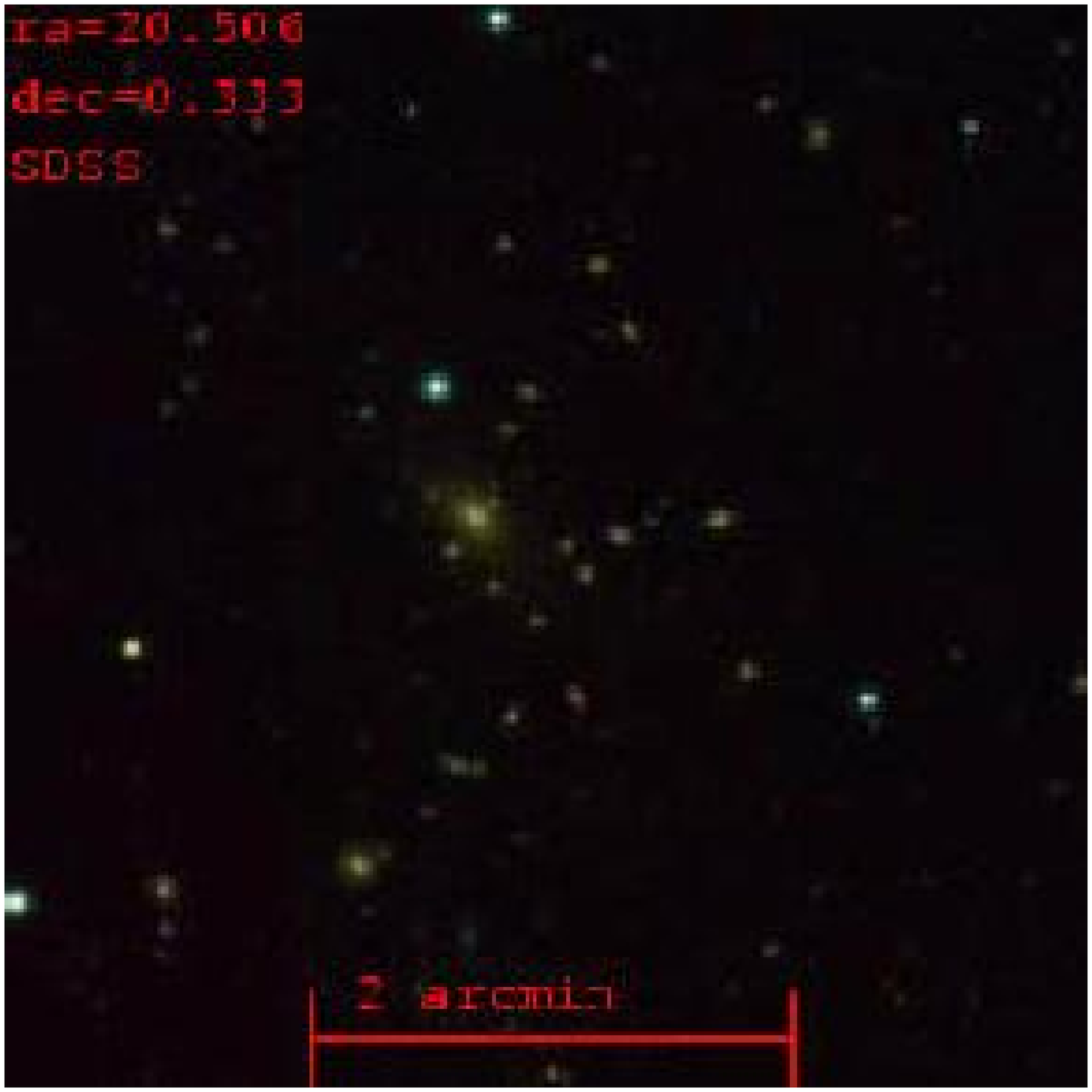}
\hspace{0.3in}
\includegraphics[width=3.1in]{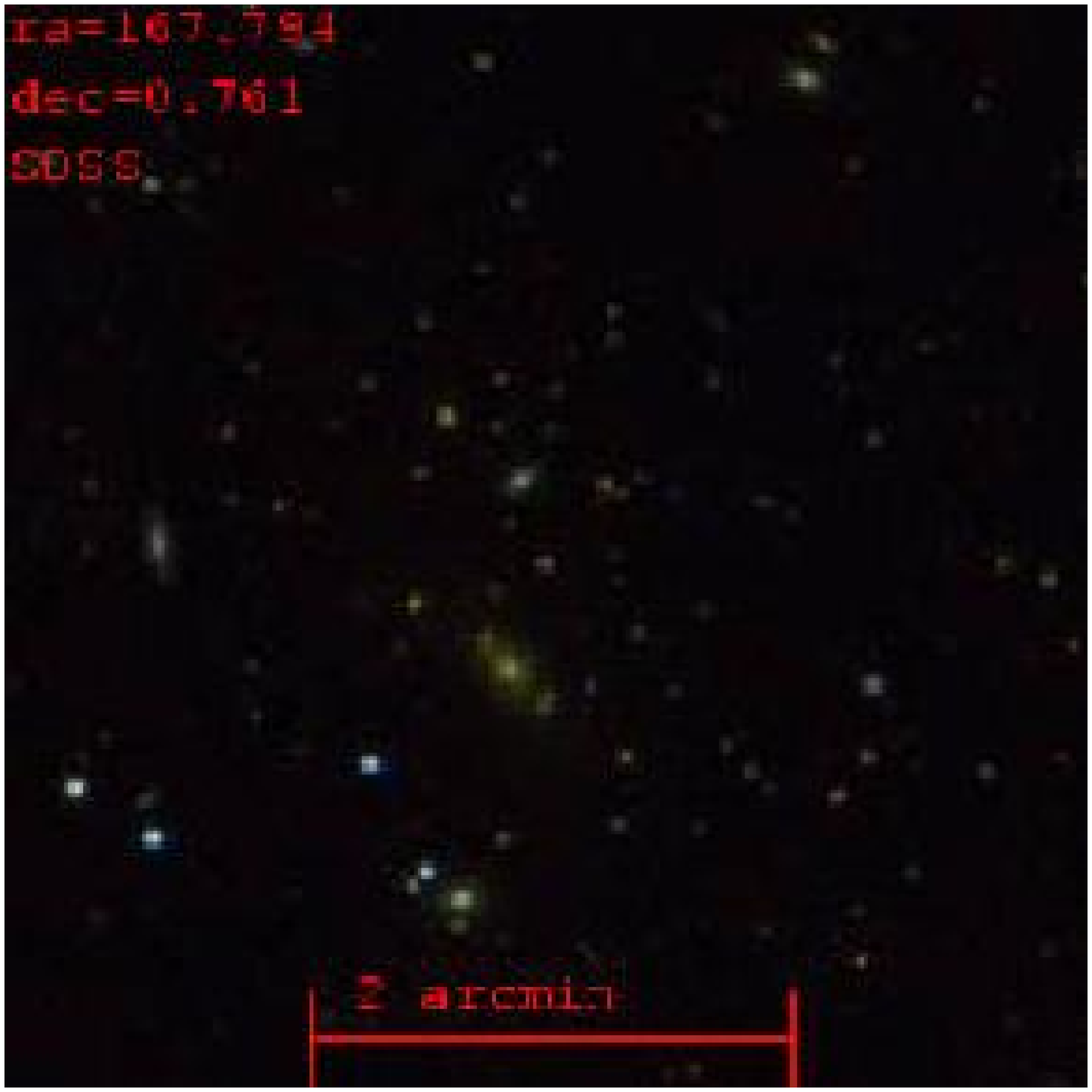}
\\
{\scriptsize BH152 (z=0.175; H,B; \lam=46.7; \ngal=32; A181; RXC0121.9)}  
\hskip 0.29in 
{\scriptsize BH379 (z=0.185; H,B; \lam=76.0; \ngal=27; A1191)}
\vskip 0.2in
\addtocounter{figure}{-1}
\caption{Continued}
\end{figure*}
\begin{figure*}
\includegraphics[width=3.1in]{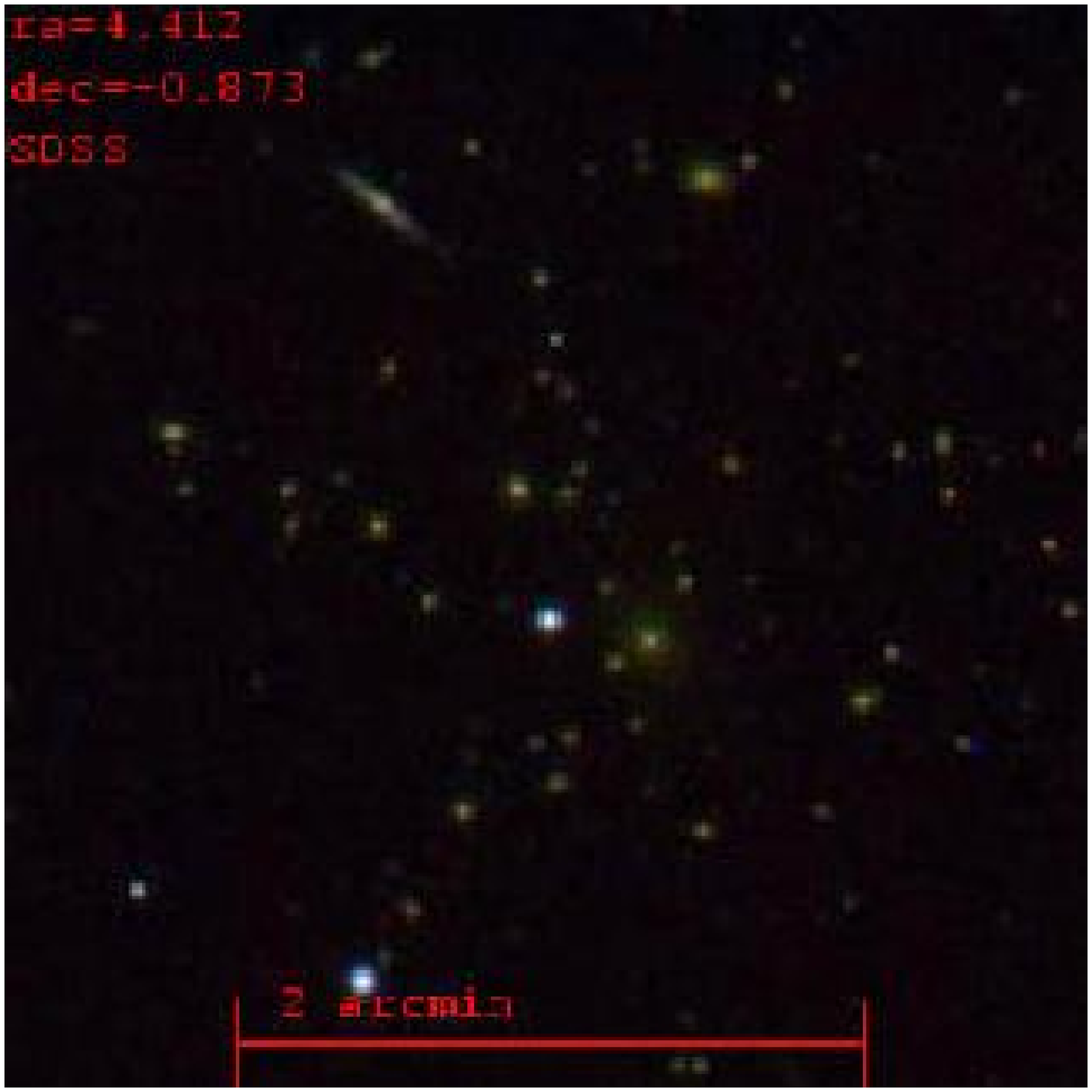}
\hspace{0.3in}
\includegraphics[width=3.1in]{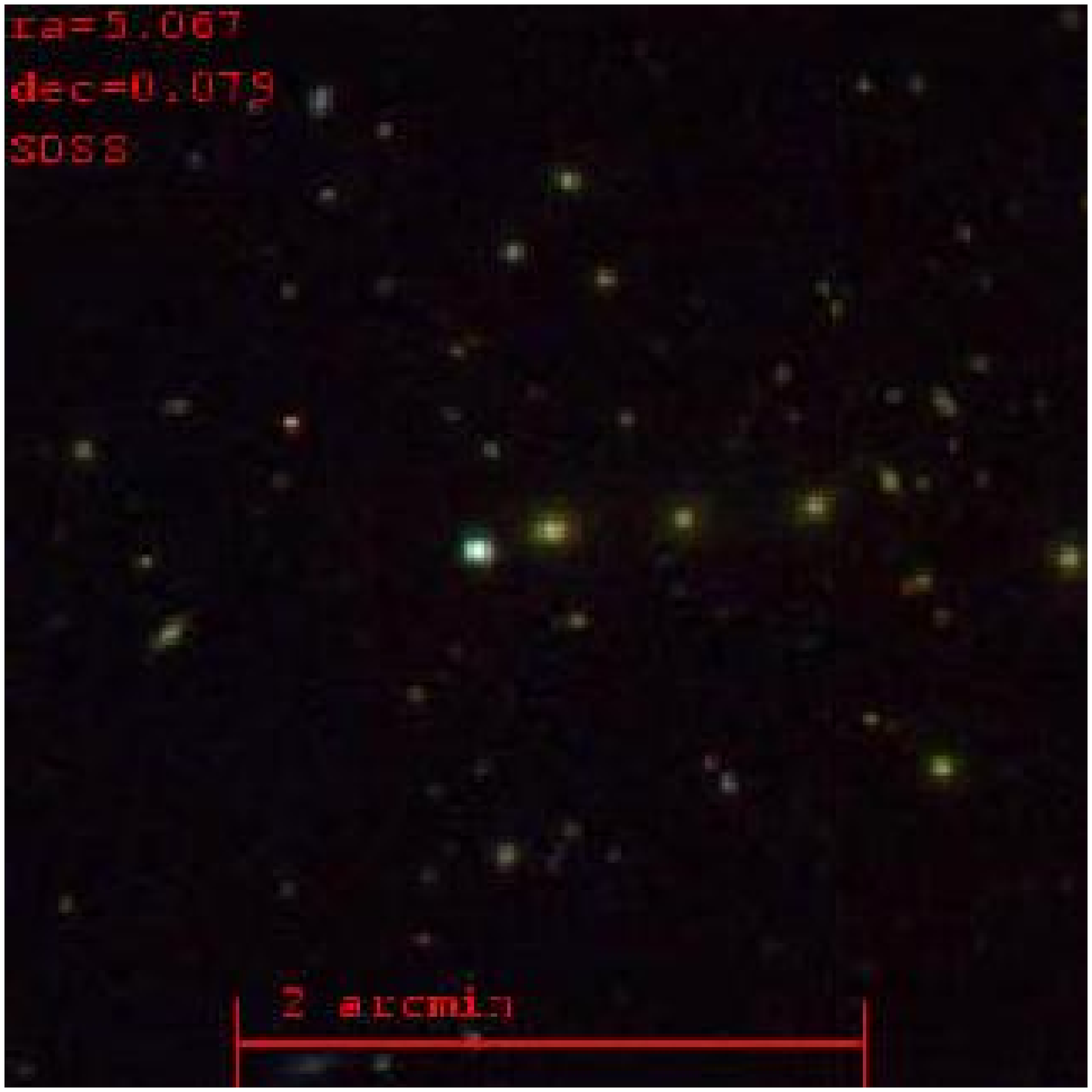}
\\
{\scriptsize BH35 (z=0.210; H,B; \lam=62.2; \ngal=26)}  
\hskip 1.34in 
{\scriptsize BH40 (z=0.210; B; \ngal=33; RXC0020.1)}
\vskip 0.2in
\includegraphics[width=3.1in]{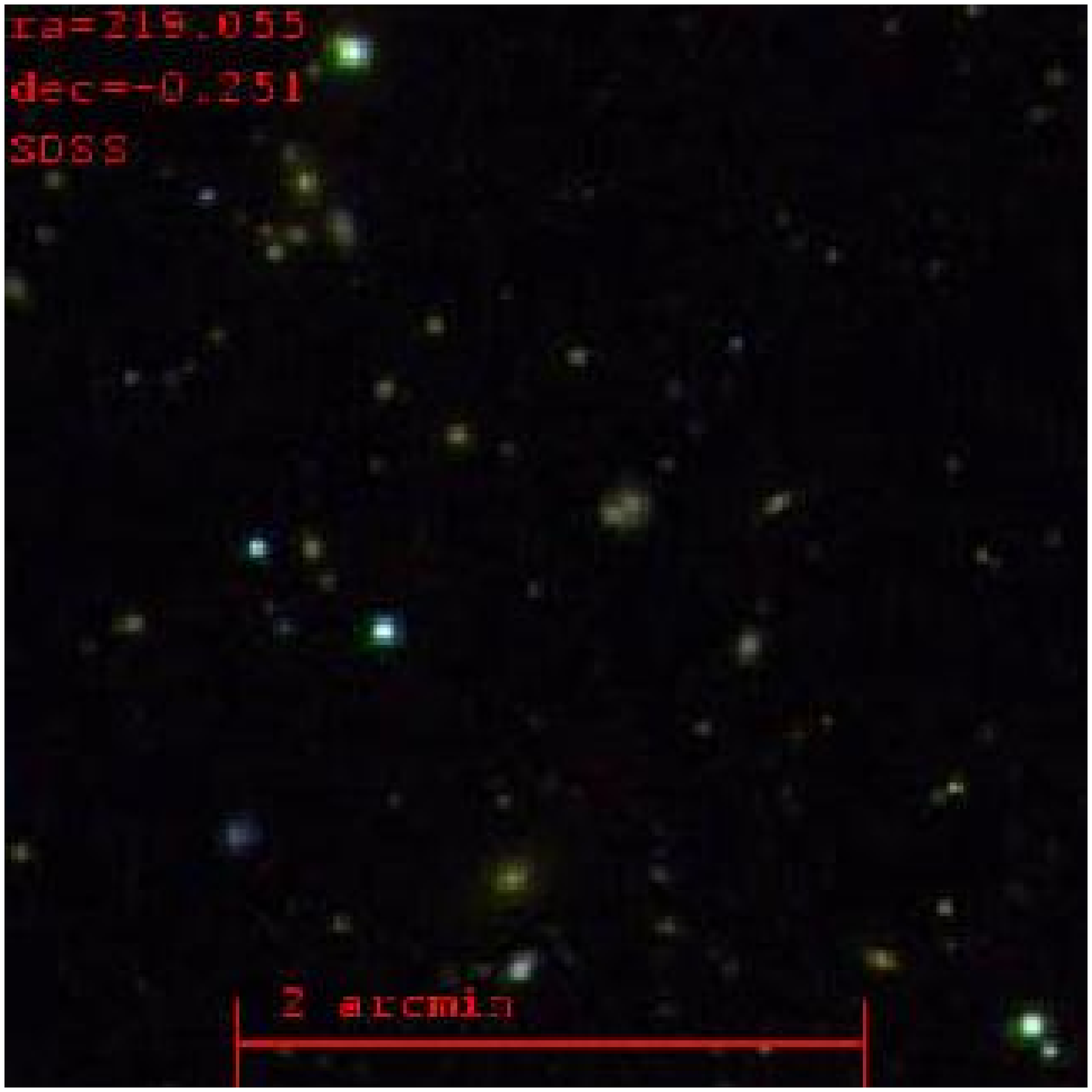}
\hspace{0.3in}
\includegraphics[width=3.1in]{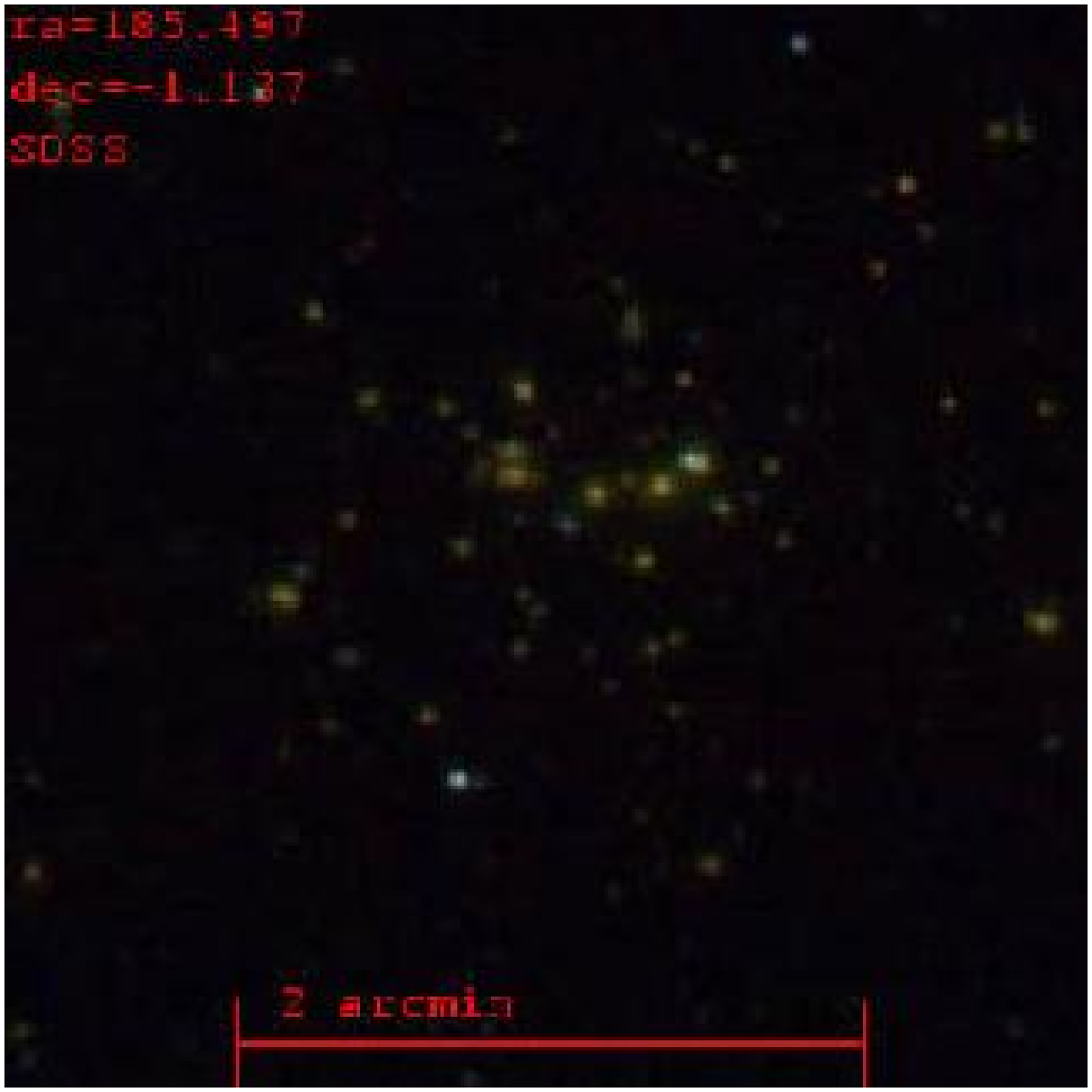}
\\
{\scriptsize BH695 (z=0.219; H,B; \lam=66.2; \ngal=16)}  
\hskip 1.28in 
{\scriptsize BH481 (z=0.225; H,B; \lam=79.0; \ngal=22; A1525)}
\vskip 0.2in
\addtocounter{figure}{-1}
\caption{Continued}
\end{figure*}
\begin{figure*}
\includegraphics[width=3.1in]{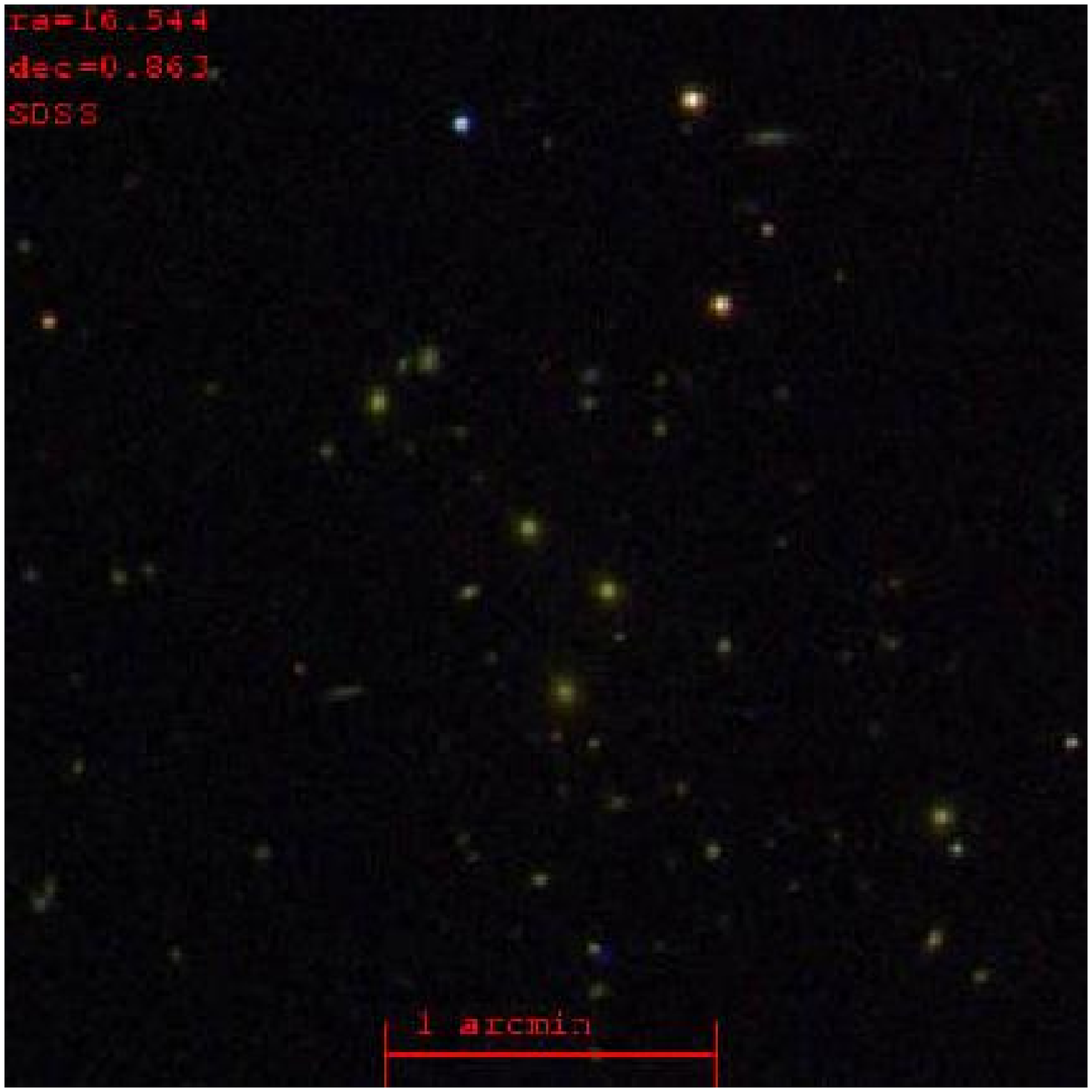}
\hspace{0.3in}
\includegraphics[width=3.1in]{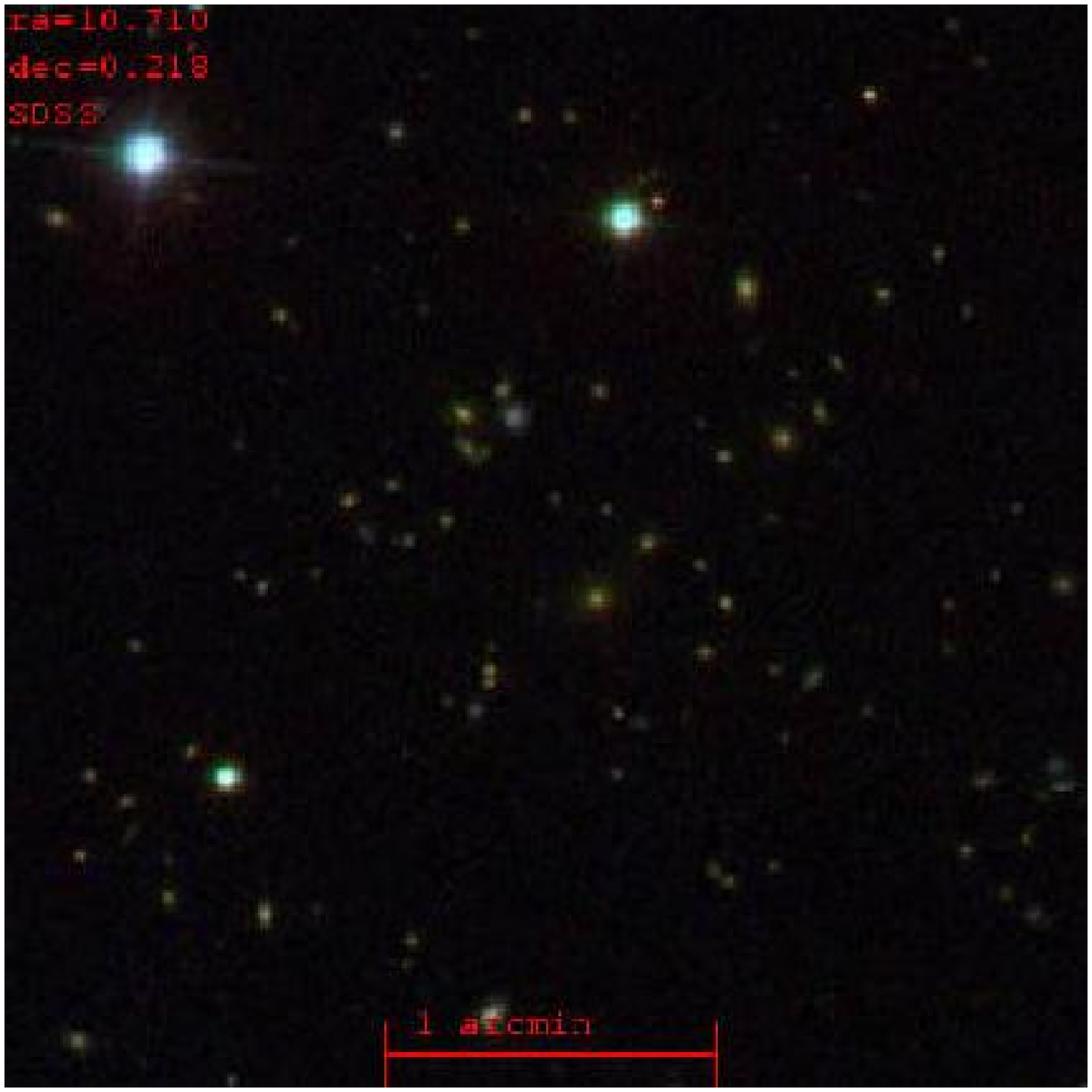}
\\
{\scriptsize BH126 (z=0.264; H,B; \lam=61.8; \ngal=22; A142)}  
\hskip 0.94in 
{\scriptsize BH94 (z=0.268; H,B; \lam=69.0; \ngal=16)}
\vskip 0.2in
\includegraphics[width=3.1in]{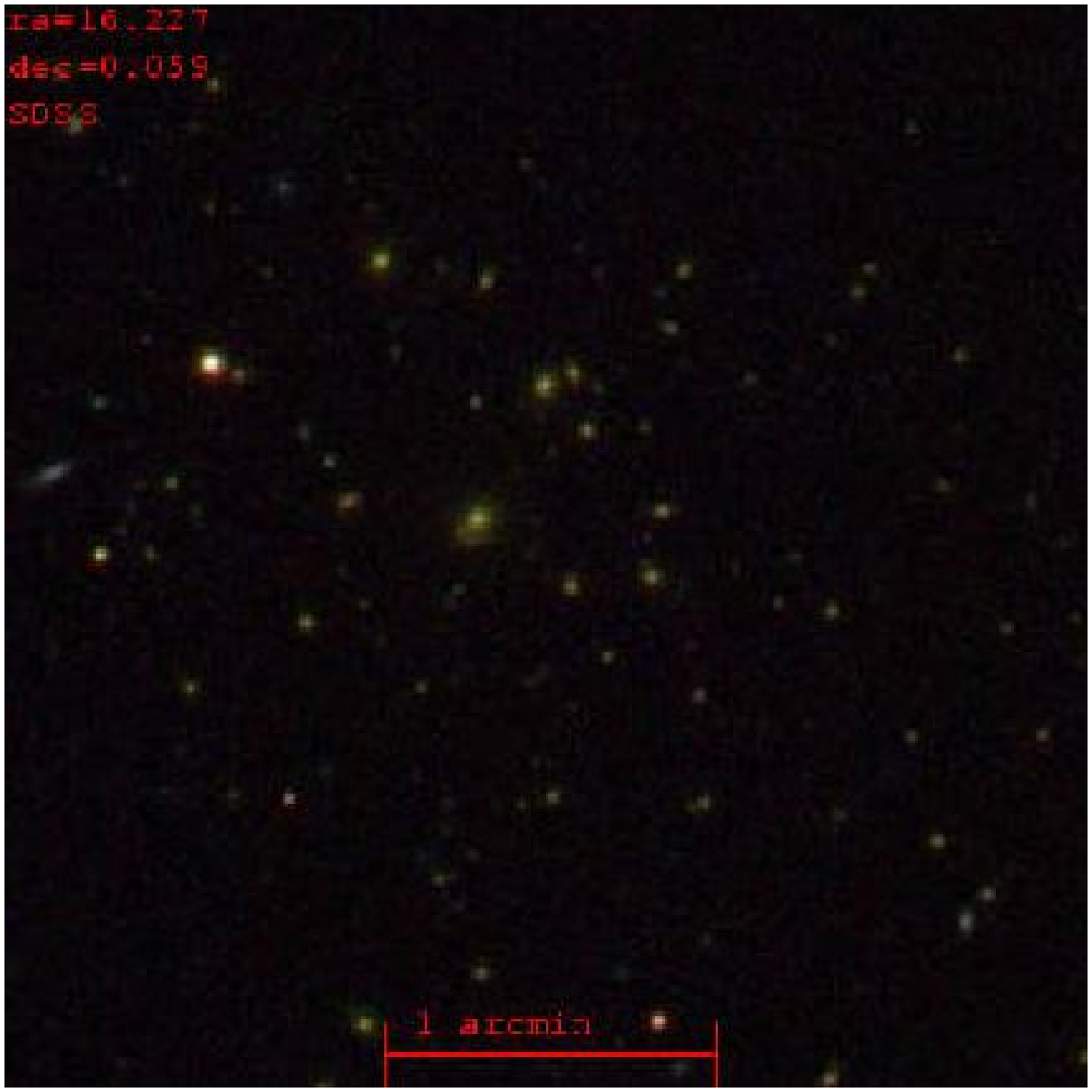}
\hspace{0.3in}
\includegraphics[width=3.1in]{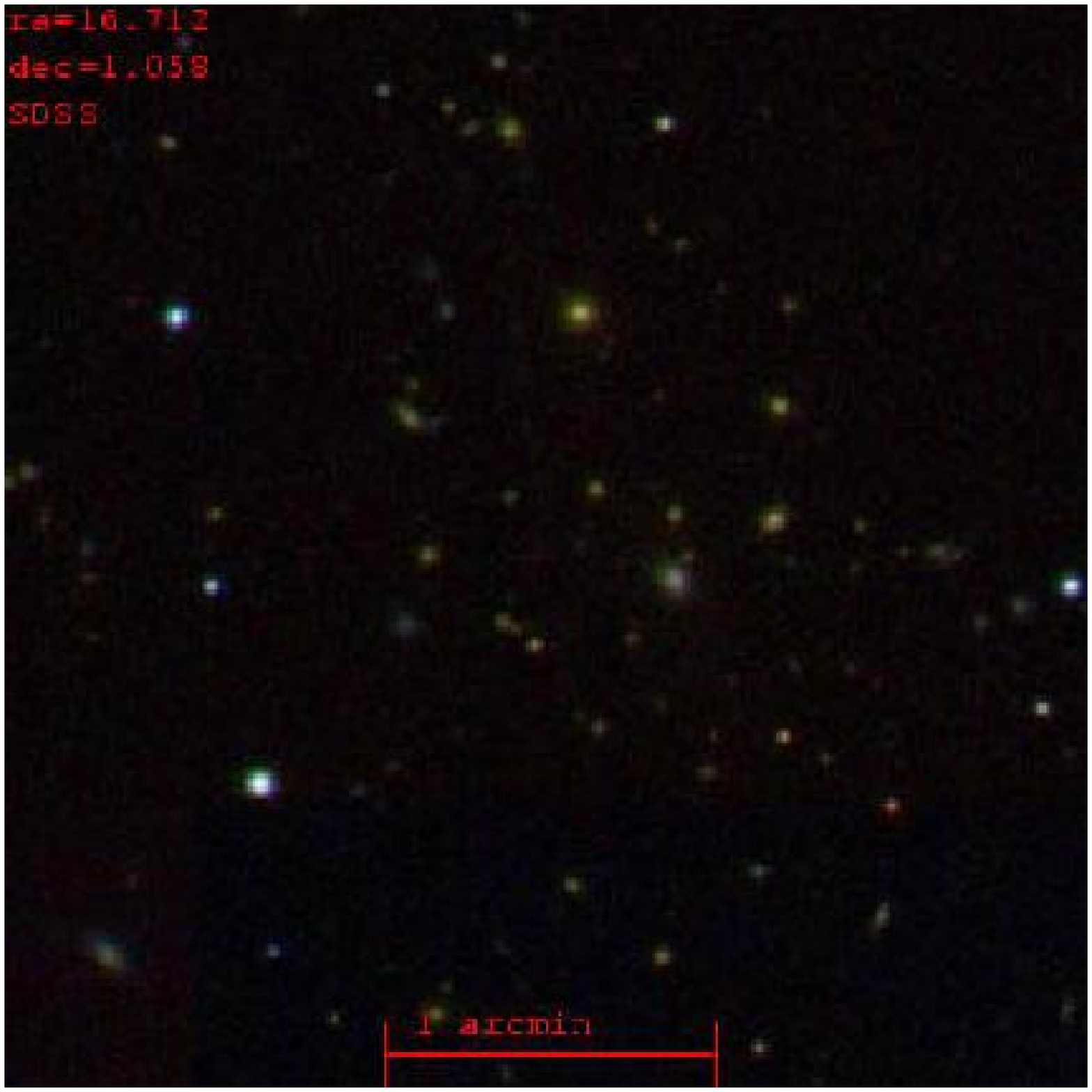}
\\
{\scriptsize BH119 (z=0.276; H,b; \lam=77.4; \ngal=35)}
\hskip 1.30in 
{\scriptsize BH131 (z=0.286; H,B; \lam=50.9; \ngal=25; RXC0106.8)}
\vskip 0.2in
\addtocounter{figure}{-1}
\caption{Continued}
\end{figure*}

\epsscale{0.95}
\begin{figure}
\plotone{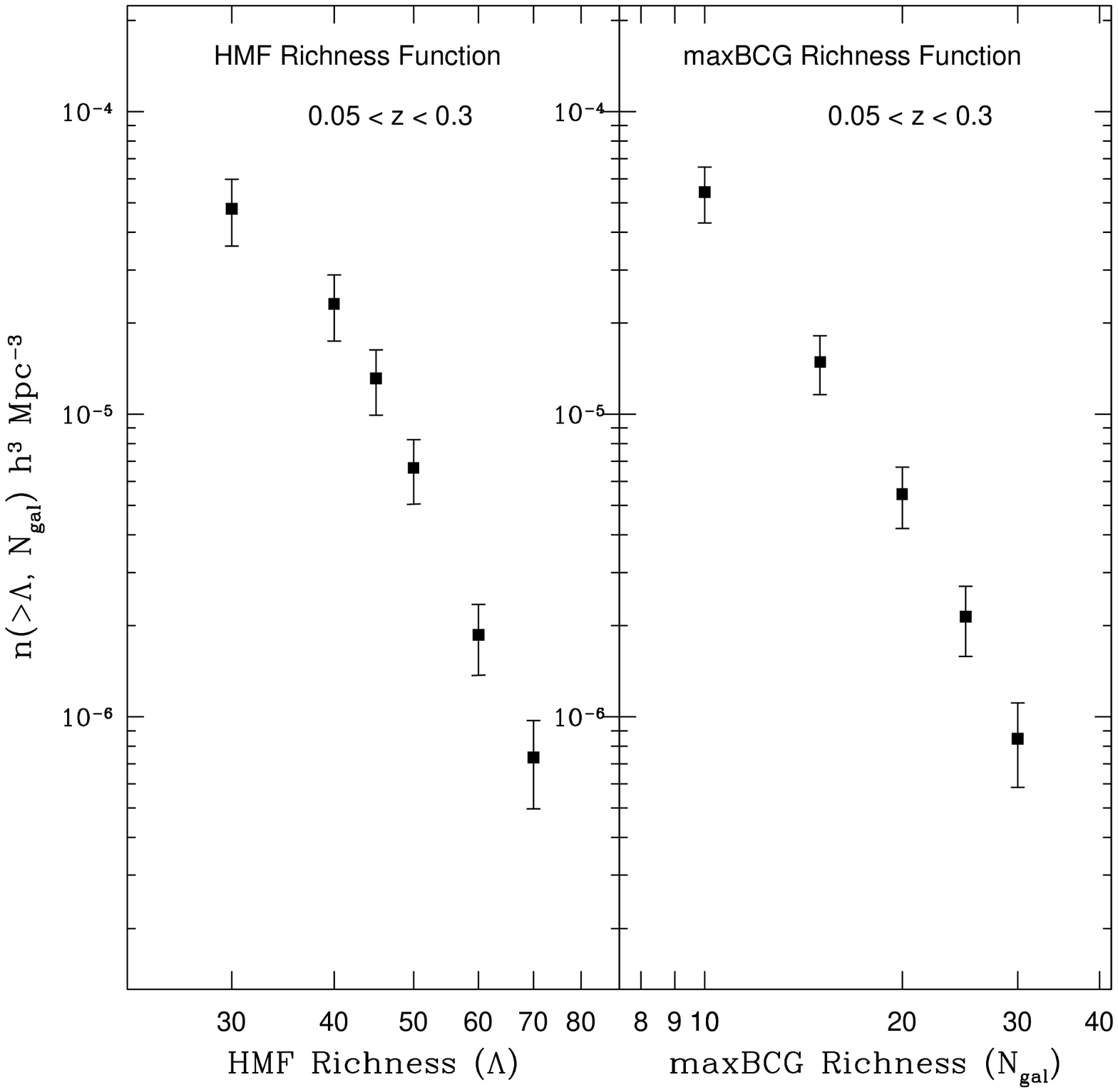}
\caption{The richness function of HMF and maxBCG clusters. The function
represents the abundance of z$_{est}$ = 0.05 - 0.3 clusters above a given richness
as a function of richness. The observed number of clusters has been corrected
by the relevant selection function and the false-positive correction factor
for each method. A flat cosmology with \om = 0.3 is used for the volume 
determination. 
\label{f14}}
\end{figure}

\clearpage

\begin{deluxetable}{rrrrr}
\tablecolumns{5} 
\tablecaption{Cluster Mean Scaling Relations}
\tablewidth{0pc}
\tablehead{
\colhead{\lam \ } & \colhead{\ngal \ \ \ }   & \colhead{\ $\sigma_v$}   &
\colhead{$L^{{}^{tot}}_{0.6}$} &
\colhead{$M_{0.6}$} \\
\colhead{} & \colhead{} & \colhead{\ ($km\ s^{-1}$)} &
\colhead{($10^{11} h^{-2} L_{\odot}$)} & 
\colhead{($10^{14} h^{-1} M_{\odot}$)} 
}
\startdata
30 \ \ &  8 \ \ \ & 300 \ \ \ \  & 1.6 \ \ \ \ \ \ \ & 0.3 \ \ \ \ \ \ \ \  \\
40 \ \ & 13 \ \ \ & 400 \ \ \ \  & 2.8 \ \ \ \ \ \ \ & 0.5 \ \ \ \ \ \ \ \  \\
50 \ \ & 20 \ \ \ & 500 \ \ \ \  & 4.4 \ \ \ \ \ \ \ & 0.9 \ \ \ \ \ \ \ \  \\
60 \ \ & 29 \ \ \ & 610 \ \ \ \  & 6.3 \ \ \ \ \ \ \ & 1.4 \ \ \ \ \ \ \ \  \\
70 \ \ & 40 \ \ \ & 720 \ \ \ \  & 8.5 \ \ \ \ \ \ \ & 2.0 \ \ \ \ \ \ \ \  \\
\cutinhead{}
\ngal \ \ & \lam \ \ \ \ & $\sigma_v$ \ \ \ \ \ & $L^{{}^{tot}}_{0.6}$ \ \ \ \ \ \ &
$M_{0.6}$ \ \ \ \ \ \ \ \\
\ \ \ & \ \ \ & ($km\ s^{-1}$) & ($10^{11} h^{-2} L_{\odot}$) \ &
($10^{14} h^{-1} M_{\odot}$) \ \\
\hline \\
10 \ \ & 35 \ \ \ & 350\ \ \ \  & 2.1 \ \ \ \ \ \ \ & 0.35  \ \ \ \ \ \ \  \\
15 \ \ & 43 \ \ \ & 430\ \ \ \  & 3.2 \ \ \ \ \ \ \ & 0.6 \ \ \ \ \ \ \ \  \\
20 \ \ & 50 \ \ \ & 500\ \ \ \  & 4.4 \ \ \ \ \ \ \ & 0.9 \ \ \ \ \ \ \ \  \\
25 \ \ & 56 \ \ \ & 560\ \ \ \  & 5.5 \ \ \ \ \ \ \ & 1.2 \ \ \ \ \ \ \ \  \\
30 \ \ & 61 \ \ \ & 620\ \ \ \  & 6.6 \ \ \ \ \ \ \ & 1.5 \ \ \ \ \ \ \ \  \\
40 \ \ & 70 \ \ \ & 730\ \ \ \  & 8.7 \ \ \ \ \ \ \ & 2.2 \ \ \ \ \ \ \ \  \\

\enddata


\end{deluxetable}

\clearpage

\begin{deluxetable}{lcrrrrrrrrrrr}
\tabletypesize{\scriptsize}
\tablecolumns{13} 
\tablecaption{BH SDSS Cluster Catalog\tablenotemark{*} \label{tbl-2}} 
\tablewidth{0pt}
\tablehead{ 
\colhead{BH \ \ } & \colhead{Method}   & 
\colhead{$\alpha$(H)} & \colhead{$\delta$(H)}   & \colhead{z$_{hmf}$} &
\colhead{\lam} & \colhead{} &\colhead{$\alpha$(B)} & \colhead{$\delta$(B)} &
\colhead{z$_{bcg}$} & \colhead{\ngal} & \colhead{z$_{spec}$} & \colhead{Comments} \\
\colhead{\# \ \ } & \colhead{} & 
\colhead{(2000)} & \colhead{(2000)} & \colhead{} &
\colhead{} & \colhead{} & \colhead{(2000)} & \colhead{(2000)} &
\colhead{} & \colhead{} & \colhead{} &
\colhead{}
}
\startdata 
    1 &H \ \ \ \             \  &  355.271 &  0.088 \  \  &  0.09 \  &  45.1 & \  &    0.000 &  0.000 \  \  &  0.000 &   0 \  \  &  0.000 & \ A2644, (RXC 2341.1)   \\  
    2 &  \ \    B            \  &    0.000 &  0.000 \  \  &  0.00 \  &   0.0 & \  &  355.277 & -0.002 \  \  &  0.205 &  16 \  \  &  0.190 & \  \\  
    3 &  \ \    B            \  &    0.000 &  0.000 \  \  &  0.00 \  &   0.0 & \  &  355.342 &  0.333 \  \  &  0.244 &  16 \  \  &  0.278 & \  \\  
    4 &  \ \    B            \  &    0.000 &  0.000 \  \  &  0.00 \  &   0.0 & \  &  355.916 &  0.424 \  \  &  0.190 &  13 \  \  &  0.186 & \  \\  
    5 &             H,B \    \  &  355.945 &  0.263 \  \  &  0.24 \  &  64.4 & \  &  355.899 &  0.331 \  \  &  0.268 &  38 \  \  &  0.000 & \  \\  
    6 &             H,B \    \  &  356.518 & -0.185 \  \  &  0.25 \  &  46.0 & \  &  356.520 & -0.186 \  \  &  0.257 &  17 \  \  &  0.266 & \  \\  
    7 &             H,B \    \  &  356.604 &  0.732 \  \  &  0.21 \  &  49.1 & \  &  356.607 &  0.728 \  \  &  0.218 &  20 \  \  &  0.000 & \  \\  
    8 &  \ \    B            \  &    0.000 &  0.000 \  \  &  0.00 \  &   0.0 & \  &  356.847 & -0.164 \  \  &  0.227 &  19 \  \  &  0.000 & \  \\
\enddata
\end{deluxetable}

\clearpage

\begin{deluxetable}{rrrrr}
\tablecolumns{5} 
\tablecaption{Cluster Catalog Statistics \label{tbl-3}}
\tablewidth{0pc}
\tablehead{ 
\colhead{\lam} & \colhead{N$_{cl}$(HMF)\tablenotemark{*}}   & 
\colhead{\ $\bigl|$} & \colhead{\ngal}   & \colhead{N$_{cl}$(maxBCG)\tablenotemark{*}} \\
\colhead{} & \colhead{(z=0.05-0.3)} & 
\colhead{\ $\bigl|$} & \colhead{}   & \colhead{(z=0.05-0.3)}
}
\startdata 
40-50 \ \ \ \ \ \ & 297 \ \ \ \ \ & \ $\bigl|$ & 13-20 \ \ \ \ \ \ \ & 402 \ \ \ \ \ \ \\
50-60 \ \ \ \ \ \ &  97 \ \ \ \ \ & \ $\bigl|$ & 20-30 \ \ \ \ \ \ \ & 103 \ \ \ \ \ \ \\
60-70 \ \ \ \ \ \ &  25 \ \ \ \ \ & \ $\bigl|$ & 30-40 \ \ \ \ \ \ \ &  15 \ \ \ \ \ \ \\
$\geq$70 \ \ \ \ \ \ &  17 \ \ \ \ \ & \ $\bigl|$ & $\geq$40 \ \ \ \ \ \ \ & 4 \ \ \ \ \ \ \\
Total(\lam$\geq$40) \ & 436 \ \ \ \ \ & \ $\bigl|$ & Total(\ngal$\geq$13) & 524 \ \ \ \ \ \ \\
\cutinhead{}
z$_{hmf}$ \ \ \ \ \ \ \ & N$_{cl}$(HMF)\tablenotemark{*} & \ $\bigl|$ & 
z$_{bcg}$ \ \ \ \ \ \ \ & N$_{cl}$(maxBCG)\tablenotemark{*} \\
 \ & (\lam$\geq$40) \ \ & \ $\bigl|$ & \ & (\ngal$\geq$13) \ \ \\
\hline \\
0.05-0.1 \ \ \ \ \ \ & 45 \ \ \ \ \ & \ $\bigl|$ & 0.05-0.1 \ \ \ \ \ & 52 \ \ \ \ \ \ \\
0.1-0.2 \ \ \ \ \ \ & 150 \ \ \ \ \ & \ $\bigl|$ & 0.1-0.2 \ \ \ \ \ & 173 \ \ \ \ \ \ \\
0.2-0.3 \ \ \ \ \ \ & 241 \ \ \ \ \ & \ $\bigl|$ & 0.2-0.3 \ \ \ \ \ & 299 \ \ \ \ \ \ \\
Total(z=0.05-0.3) & 436 \ \ \ \ \ & \ $\bigl|$ & Total(z=0.05-0.3) & 524 \ \ \ \ \ \ \\

\enddata


\tablenotetext{*}{N$_{cl}$ is the number of clusters observed, uncorrected 
for selection function and false-positive detections. Survey area = 379 deg$^2$.}

\end{deluxetable}

\clearpage

\begin{deluxetable}{rrrrrrrrr}
\tablecolumns{9} 
\tablecaption{Comparison to NORAS X-ray clusters \label{tbl-4}}
\tablewidth{0pc}
\tablehead{ 
\colhead{NORAS cluster} & \colhead{$\alpha$(2000)}   & 
\colhead{$\delta$(2000)} & \colhead{z}   & \colhead{$L_x^{44}$} &
\colhead{z$_{bcg}$} & \colhead{\ngal} & 
\colhead{z$_{hmf}$} & \colhead{\lam} 
}
\startdata 
RXC J0020.1+0005 & 5.044 & 0.092 \ & 0.212 & 1.08 &
0.220  & 33 \ & --- \ & --- \ \\
RXC J0106.8+0103 & 16.709 & 1.055 \ & 0.253 & 8.65 &
0.286  & 25 \ & 0.22 & 50.9 \\
RXC J0114.9+0024 & 18.740 & 0.406 \ & 0.044 & 0.92 &
0.068  & 12 \ & 0.05 & 42.6 \\
RXC J0121.9+0021 & 20.493 & 0.358 \ & 0.175 & 1.85 &
0.166  & 32 \ & 0.16 & 46.7 \\
RXC J0152.7+0100 & 28.182 & 1.016 \ & 0.227 & 9.37 &
0.264  & 29 \ & 0.23 & 73.4 \\
RXC J1326.2+0013 & 201.573 & 0.225 \ & 0.082 & 1.65 &
0.098  & 13 \ & 0.13 & 44.1 \\
RXC J1415.8+0015 & 213.965 & 0.259 \ & 0.125 & 0.83 &
0.129  & 15 \ & 0.13 & 49.4 \\
RXC J2341.1+0018 & 355.276 & 0.315 \ & 0.110 & 1.03 &
0.140  &  8 \ & 0.09 & 45.1 \\

\enddata


\end{deluxetable}

\end{document}